\documentclass[amsmath,amssymb,aps,pra,superscriptaddress,twocolumn
]{revtex4-2}

\usepackage{amssymb}
\usepackage{graphicx}
\usepackage{dcolumn}
\usepackage{bm}
\usepackage{amsmath}
\usepackage[colorlinks,urlcolor=cyan,citecolor=blue,linkcolor=magenta]{hyperref}

\begin{abstract}
We study the problem of two harmonically trapped atoms in the presence of spin-orbital-angular-momentum (SOAM) coupling. The two-body energy spectrum is numerically calculated by utilizing the exact diagonalization method. We analyze how the degeneracy of energy levels is lifted under the interplay between the interatomic interaction and SOAM coupling. The exact numerical results show excellent agreement with that of perturbation theory in the weak-interaction limit as well as that in the absence of SOAM coupling. The properties of correlations between the two atoms are also discussed with respect to the interaction strength. The findings in this paper may provide valuable insights into few-body physics subjected to SOAM coupling and the possible experimental detection of spectrum functions in many-body systems, such as radiofrequency spectroscopy. 
\end{abstract}

\begin{document}
\title{Two atoms in a harmonic trap with spin-orbital-angular-momentum coupling}
\author{Xiao-Long Chen}
\affiliation{Department of Physics, Zhejiang Sci-Tech University, Hangzhou 310018, China}
\affiliation{Zhejiang Key Laboratory of Quantum State Control and Optical Field Manipulation, Zhejiang Sci-Tech University, Hangzhou 310018, China}

\author{Aixi Chen}
\affiliation{Department of Physics, Zhejiang Sci-Tech University, Hangzhou 310018, China}
\affiliation{Zhejiang Key Laboratory of Quantum State Control and Optical Field Manipulation, Zhejiang Sci-Tech University, Hangzhou 310018, China}

\author{Shi-Guo Peng}
\email{pengshiguo@wipm.ac.cn}
\affiliation{State Key Laboratory of Magnetic Resonance and Atomic and Molecular
Physics, Innovation Academy for Precision Measurement Science and Technology, Chinese Academy of Sciences, Wuhan 430071, China}
\date{\today}

\maketitle

\section{Introduction}

The coupling between the orbital angular momentum (OAM) $\bf{L}$ of a charged particle and its spin $\bf{S}$, known as the $LS$ coupling, plays a fundamental role in few-body problems across diverse fields of physics. In atomic physics, $LS$ coupling contributes to the fine structure of atomic spectra and explains the splitting of spectra lines into multiple components in the presence of an external magnetic field~\cite{Landau2007Q}. In nuclear physics, $LS$ coupling provides a dominant interaction mechanism in understanding the nuclear structure in the framework of the shell model~\cite{de2013nuclear}. Taking a broader perspective, the spin-orbit (SO) coupling or interaction lies at the heart of several many-body phenomena in condensed matter physics, by influencing the band structure and electronic transport properties and giving rise to fascinating emergent phenomena such as the spin Hall effect and topological insulators~\cite{Qi2010T,hasan2010C}. Hence, understanding and controlling SO coupling in materials are essential for advancing technologies such as spintronics, topological materials, and quantum computation.

In recent decades, the successful realization of SO coupling in cold bosonic and fermionic atoms has provided a remarkably flexible playground to study these fascinating phenomena closely associated with SO coupling in a highly controllable way~\cite{lin2011spin,wang2012spin,cheuk2012spin}. Though the SO coupling has been intensively studied both experimentally and theoretically in cold atoms during past years ~\cite{liu2009effect,lin2011spin,williams2012synthetic,wang2012spin,cheuk2012spin,zhang2012collective,olson2014tunable,fu2014production,ji2014experimental,ji2015softening,hamner2015spin,garcia2015tunable,burdick2016long,song2016spin,li2016spin,livi2016synthetic,osterloh2005cold,ruseckas2005non,juzeliunas2010generalized,campbell2011realistic,sau2011chiral,anderson2013magnetically,xu2013atomic,liu2014realization,anderson2012synthetic,lu2020ideal,wang2018dirac,huang2016experimental,wu2016realization,wang2021realization,galitski2013spin,zhai2015degenerate,Zhang2018S}, it was not until very recently that the type of $LS$ coupling, or say spin-OAM (SOAM) coupling~\cite{Liu2006G,demarco2015angular,sun2015spin,qu2015quantum,hu2015half,chen2016spin,Vasic2016E,Hou2017A}, was achieved in experiments with cold atoms~\cite{chen2018spin,chen2018rotating,zhang2019ground} and has stimulated fruitful studies of such an intriguing quantum system~\cite{peng2022spin,chen2020angular,chen2020generating,wang2021exotic,duan2020symmetry,bidasyuk2022fine,chen2022angular,cao2022quantum,han2022molecular}. Unlike the situation in condensed matter physics that the SOAM coupling of electrons is a relativistic effect and much weaker than the Coulomb interaction, the energy scale of SOAM coupling realized in cold atomic experiments could be comparable with that of interatomic interactions and even to the many-body characteristic energy scale. This challenges the conventional perturbation theory for calculating the few-body energy spectrum in dealing with SOAM coupling. As a consequence, the difficulty may lie, for example, in the two-body problem that the separation between the center of mass (c.m.) and relative motions is not straightforward even in the Hamiltonian, posing significant challenges for theoretical treatment. 

In this paper, we theoretically study the problem of two harmonically trapped atoms in the presence of the type of SOAM coupling achieved in recent experiments with $^{87}$Rb atomic gases~\cite{chen2018spin,chen2018rotating,zhang2019ground}. An unperturbed theoretical framework is developed for solving the two-body Schr\"{o}dinger equation. By considering a tunable interaction potential and the SOAM coupling, the energy spectra and eigenfunctions of two atoms in a harmonic trap are numerically calculated by solving the derived secular equation via the approach of exact diagonalization. We demonstrate the intrinsic mechanism underlying the elimination of degeneracy in two-body energy levels due to the interplay between the interatomic interaction and SOAM coupling. Our numerical results show excellent consistency with that in the limit without SOAM coupling as well as that of perturbation theory in the weak-interaction limit. We further introduce a correlation function and find that the correlations between the two atoms are significantly modified by SOAM coupling. 

The rest of this paper is organized as follows. The model single- and two-body Hamiltonians are introduced in Sec.~\ref{sec: model}. In Sec.~\ref{sec:single-body}, we introduce the single-body problem and show the expressions of the single-body wavefunctions as well as the eigenenergies. In Sec.~\ref{sec:two-body}, we further discuss the two-body problem and develop a secular equation for the eigenenergy and eigenfunction in the presence of SOAM coupling. Finally, in Sec.~\ref{sec: results}, our numerical results and findings are presented. We first justify the numerically calculated energy spectrum without SOAM coupling by comparing it with the analytic result. We then illustrate the energy spectrum as well as an introduced correlation function in the presence of SOAM coupling and discuss the role of interaction potential and SOAM coupling. A summary is given in Sec.~\ref{sec: summary}.

\section{Hamiltonian} \label{sec: model}

\begin{figure}[t]
\includegraphics[width=0.48\textwidth]{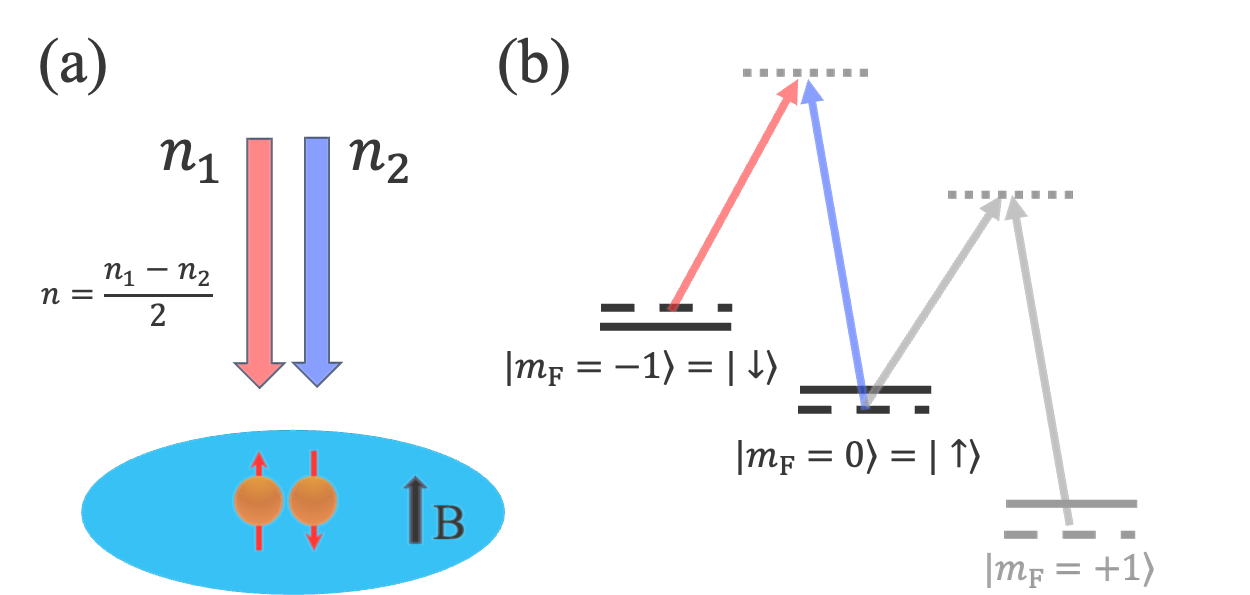}
\caption{(Color online) Illustration of the experimental scheme of spin-orbital-angular-momentum (SOAM) coupling using cold atoms. (a) Two Laguerre-Gaussian laser beams with orbital angular momentum (OAM) $n_{1,2}\hbar$ couple the cold atomic cloud with a biased magnetic field $B$ to induce the hyperfine structure. The induced OAM transferred to the cloud is $n\hbar=(n_1-n_2)\hbar/2$. (b) Two hyperfine states of $^{87}$Rb are involved by Raman beams, which play the roles of pseudospins $\left|\uparrow\right>\equiv\left|F=1,m_F=0\right>$ and $\left|\downarrow\right>\equiv\left|F=1,m_F=-1\right>$, equivalent to an effective spin-half system.}
\label{figure1_scheme}
\end{figure}
The SOAM-coupling effect has been achieved in bosonic $^{87}$Rb atoms by using a pair of copropagating Raman beams operated in Laguerre-Gaussian (LG) modes with opposite angular momenta \cite{chen2018spin,chen2018rotating,zhang2019ground}. Explicitly, as illustrated in Fig.~\ref{figure1_scheme}, two LG laser beams carrying OAM $n_{1,2}\hbar$ couple the internal spin states of cold rubidium atoms split by a biased magnetic field $B$, and the system can then be described by an effective two-level scheme with the third spin state far away from other two. The {\it external} OAM of atoms changes by $n\hbar=(n_1-n_2)\hbar/2$ when transitioning between two {\it internal} ground hyperfine states. It leads to a so-called Raman-induced SOAM coupling, which is effectively described by the Hamiltonian~\cite{zhang2019ground}:
\begin{equation}
\hat{H}=\hat{H}_\mathrm{ho}+\Omega\left(r\right)\hat{\sigma}_{x}-\frac{n\hbar}{mr^{2}}\hat{l}_{z}\hat{\sigma}_{z}+\frac{\left(n\hbar\right)^{2}}{2mr^{2}},
\end{equation}
at resonance with a vanishing two-photon detuning. Here, $\hat{H}_\mathrm{ho}=-\hbar^{2}\nabla^{2}/2m+m\omega^{2}r^{2}/2$ is the Hamiltonian of a harmonic oscillator with trapping frequency $\omega$, $\Omega\left(r\right)=\Omega_{R}\left(r/w\right)^{2\left|n\right|}\mathrm{exp}{(-2r^{2}/w^{2})}$ is the effective transverse Zeeman field with the coupling strength $\Omega_{R}$ and the waist $w$ of LG beams, and $n$ is the angular momentum transferred from LG beams to atoms during the Raman process. We have adopted the polar coordinate ${\bf r}=\left(r,\varphi\right)$, $\hat{l}_{z}=-i\hbar\partial_{\varphi}$ is the angular momentum operator, and $\hat{\sigma}_{x,z}$ are Pauli matrices. Apparently, the key feature of SOAM coupling is characterized by the term $\hat{l}_{z}\hat{\sigma}_{z}$ that is of a two-dimensional (2D) nature. Thus, we will adopt a 2D geometry that can sufficiently capture the crucial physics of SOAM coupling in such systems as in our previous works~\cite{zhang2019ground,chen2020angular}. In the experiment~\cite{zhang2019ground}, the waist of LG beams is $\sim63$ $\mu$m, much larger than the size of condensate, i.e., $r\ll w$. Thus, the effective transverse Zeeman field $\Omega\left(r\right)$ experienced by atoms is considerably weak and negligible within the length scale of the atomic cloud. As a result, the single-body Hamiltonian in the presence of SOAM coupling can be further simplified to~\cite{chen2020angular}
\begin{equation}
\hat{H}=-\frac{\hbar^{2}}{2mr}\frac{\partial}{\partial r}r\frac{\partial}{\partial r}+\frac{1}{2}m\omega^{2}r^{2}+\frac{\left(\hat{l}_{z}-n\hbar\hat{\sigma}_{z}\right)^{2}}{2mr^{2}},\label{eq:1bHamiltonian}
\end{equation}
which captures the key feature of SOAM coupling.

The two-body Hamiltonian takes the form of
\begin{equation} \label{eq:2bHamiltonian}
\hat{\mathcal{H}}=\sum_{i=1}^{2}\hat{H}_{i}+\hat{U}\left({\bf r}_{1},{\bf r}_{2}\right),
\end{equation}
where $\hat{H}_{i}$ is the single-body Hamiltonian of the $i$th atom, and $\hat{U}\left({\bf r}_{1},{\bf r}_{2}\right)$ is the two-body interaction potential. Here, we consider an interaction existing in the spin-singlet channel as conventionally considered in a spin-half system without loss of generality, i.e.,
\begin{equation}
\hat{U}\left({\bf r}_{1},{\bf r}_{2}\right)=V\left(r_{12}\right)\left|0,0\right\rangle \left\langle 0,0\right|,
\label{eq: Interaction}
\end{equation}
where we have introduced
\begin{equation}
\left|0,0\right\rangle \equiv \left|S=0,S_{z}=0\right\rangle =\frac{1}{\sqrt{2}}\left(\left|\uparrow\downarrow\right\rangle -\left|\downarrow\uparrow\right\rangle \right),
\end{equation}
with the total spin $S=0$ and the corresponding magnetic quantum number $S_{z}=0$ along the $z$ axis. Here, $V\left(r_{12}\right)$ depends only on the distance $r_{12}=\left|{\bf{r}}_{12}\right|\equiv\left|{\bf r}_{1}-{\bf r}_{2}\right|$ between atoms. Note that our approach can be straightforwardly generalized to other forms of interactions, such as spin-triplet or spin-independent interactions, which depend on the specific atomic species.

\section{Single-body problem} \label{sec:single-body}

For a single atom, it is easily found that the angular momentum $\hat{l}_{z}$ is conserved as well as the spin $\hat{\sigma}_{z}$ along the $z$ axis. Thus, the single-body problem can be solved for given angular momentum and spin. The associated single-body wave function is then written as
\begin{equation}
\left|\psi_{l\sigma}\right\rangle =u_{l\sigma}\left|l,\sigma\right\rangle, 
\end{equation}
with
\begin{equation}
\left\langle \varphi|l,\sigma\right\rangle =\frac{e^{il\varphi}}{\sqrt{2\pi}}\left|\sigma\right\rangle ,\;\left(\sigma=\uparrow,\downarrow\right).
\end{equation}
In the spatial representation, we have
\begin{equation} \label{eq:1bWaveFunction}
\psi_{l\sigma}\left({\bf r}\right)=\left\langle {\bf r}|\psi_{l\sigma}\right\rangle =u_{l\sigma}\left(r\right)\frac{e^{il\varphi}}{\sqrt{2\pi}}\left|\sigma\right\rangle .
\end{equation}
After inserting Eq.~\eqref{eq:1bWaveFunction} into the single-body Schr\"{o}dinger equation $\hat{H}\psi_{l\sigma}\left({\bf r}\right)=\varepsilon\psi_{l\sigma}\left({\bf r}\right)$, we obtain the radial equation satisfied by $u_{l\sigma}\left(r\right)$, i.e.,
\begin{equation}
\left[\frac{d^{2}}{d\xi^{2}}+\frac{1}{\xi}\frac{d}{d\xi}-\frac{\left(l\mp n\right)^{2}}{\xi^{2}}+\left(\frac{2\varepsilon}{\hbar\omega}-\xi^{2}\right)\right]u_{l\sigma}=0,
\end{equation}
with $\xi=r/a_\mathrm{ho}$ and $a_\mathrm{ho}=\sqrt{\hbar/m\omega}$, which is the standard radial equation of a 2D harmonic oscillator, but with the angular momentum number $l\mp n$ corresponding to the spin $\sigma=\uparrow,\downarrow$, respectively. The explicit form of the radial wave function $u^{k}_{l\sigma}\left(r\right)$ is
\begin{equation}
u^{k}_{l\sigma}\left(r\right)=\mathcal{N}_{kl\sigma}\left(\frac{r}{a_\mathrm{ho}}\right)^{\left|l\mp n\right|}\mathrm{exp}{\left(-\frac{r^{2}}{2a_\mathrm{ho}^{2}}\right)}L_{k}^{\left|l\mp n\right|}\left(\frac{r^{2}}{a_\mathrm{ho}^{2}}\right),
\end{equation}
and 
\begin{equation}
\mathcal{N}^{k}_{l\sigma}=\frac{1}{a_\mathrm{ho}}\sqrt{\frac{2\cdot k!}{\left(k+\left|l\mp n\right|\right)!}}
\end{equation}
is the normalization coefficient. Here, $L_{k}^{l}\left(\cdot \right)$ is the associated Laguerre polynomials. Finally, the eigenstates of the single-body problem are characterized by three quantum numbers, i.e., the principle quantum number $k$, the angular quantum number $l$, and the spin $\sigma$, i.e.,
\begin{equation}
\psi_{kl\sigma}\left({\bf r}\right)=u^{k}_{l\sigma}\left(r\right)\frac{e^{il\varphi}}{\sqrt{2\pi}}\left|\sigma\right\rangle.
\end{equation}
The corresponding eigenenergy is
\begin{equation} \label{eq:E-single-soamc}
\varepsilon_{kl\sigma}=\left(2k+\left|l\mp n\right|+1\right)\hbar\omega.
\end{equation}
Here, we recall that $-,+$ corresponds to the results of spin $\sigma=\uparrow,\downarrow$, respectively.

\section{Two-body problem} \label{sec:two-body}

Unlike the situation in the absence of SOAM coupling, the c.m. and relative motions of two atoms are coupled by SOAM coupling, even in the Hamiltonian. The angular part of a two-body state is characterized by four quantum numbers, i.e., $\left\{ l_{1z},l_{2z};s_{1z},s_{2z}\right\} $, where $l_{iz}$ and $s_{iz}$ are, respectively, the OAM and the $z$-axis spin projection for the $i$th atom. However, they are not good quantum numbers in the presence of interaction. Obviously, $\hat{U}$ does not commute with $\hat{l}_{iz}$, and it flips the spin as well, for example,
\begin{equation}
\hat{U}\left|\uparrow\downarrow\right\rangle =\frac{V\left(r_{12}\right)}{2}\left(\left|\uparrow\downarrow\right\rangle -\left|\downarrow\uparrow\right\rangle \right).
\end{equation}
Fortunately, the projection of the total spin $S$ on the $z$ axis, i.e., $S_{z}$, is conserved by the interaction. This is easily seen by expanding the two-body Hamiltonian in the spin basis $\left\{ \left|\uparrow\downarrow\right\rangle,\left|\downarrow\uparrow\right\rangle,\left|\uparrow\uparrow\right\rangle,\left|\downarrow\downarrow\right\rangle \right\}$. We find that the two-body Hamiltonian is diagonalized in three blocks corresponding to $S_{z}=0,\pm1$ respectively, i.e.,
\begin{equation}
\hat{\mathcal{H}}=
\begin{bmatrix}
\hat{H}_{\uparrow\downarrow}^{(0)}+V/2 & -V/2 & 0 & 0\\
-V/2 & \hat{H}_{\downarrow\uparrow}^{(0)}+V/2 & 0 & 0\\
0 & 0 & \hat{H}_{\uparrow\uparrow}^{(0)} &0\\
0 &0  &0  & \hat{H}_{\downarrow\downarrow}^{(0)}
\end{bmatrix},
\end{equation}
where we have
\begin{subequations}
\begin{eqnarray}
\hat{H}_{\uparrow\downarrow}^{(0)} & = & \sum_{i}\hat{H}_{i,r}+\frac{\left(\hat{l}_{1z}-n\hbar\right)^{2}}{2mr_{1}^{2}}+\frac{\left(\hat{l}_{2z}+n\hbar\right)^{2}}{2mr_{2}^{2}},\\
\hat{H}_{\downarrow\uparrow}^{(0)} & = & \sum_{i}\hat{H}_{i,r}+\frac{\left(\hat{l}_{1z}+n\hbar\right)^{2}}{2mr_{1}^{2}}+\frac{\left(\hat{l}_{2z}-n\hbar\right)^{2}}{2mr_{2}^{2}},\\
\hat{H}_{\uparrow\uparrow}^{(0)} & = & \sum_{i}\hat{H}_{i,r}+\frac{\left(\hat{l}_{1z}-n\hbar\right)^{2}}{2mr_{1}^{2}}+\frac{\left(\hat{l}_{2z}-n\hbar\right)^{2}}{2mr_{2}^{2}},\\
\hat{H}_{\downarrow\downarrow}^{(0)} & = & \sum_{i}\hat{H}_{i,r}+\frac{\left(\hat{l}_{1z}+n\hbar\right)^{2}}{2mr_{1}^{2}}+\frac{\left(\hat{l}_{2z}+n\hbar\right)^{2}}{2mr_{2}^{2}},
\end{eqnarray}
\end{subequations}
with 
\begin{equation}
\hat{H}_{i,r}=-\frac{\hbar^{2}}{2mr_{i}}\frac{\partial}{\partial r_{i}}r_{i}\frac{\partial}{\partial r_{i}}+\frac{1}{2}m\omega^{2}r_{i}^{2}.
\end{equation}
The interaction is involved in the subspace of $S_{z}=0$, as anticipated. Therefore, it is reasonable to consider the two-body solution in the subspace of $S_{z}=0$, i.e., $\left\{\left|\uparrow\downarrow\right\rangle,\left|\downarrow\uparrow\right\rangle \right\}$, while the solutions in the subspaces of $S_{z}=\pm1$ are simply ones of free atoms. In the subspace of $S_{z}=0$, the Hamiltonian takes the explicit form of
\begin{equation} \label{eq: Hamiltonian_sub}
\hat{\mathcal{H}}=
\begin{bmatrix}
\hat{H}_{\uparrow\downarrow}^{(0)}+V\left(r_{12}\right)/2 & -V\left(r_{12}\right)/2\\
-V\left(r_{12}\right)/2 & \hat{H}_{\downarrow\uparrow}^{(0)}+V\left(r_{12}\right)/2 
\end{bmatrix}.
\end{equation}
In addition, the total OAM $\hat{L}_{z}=\hat{l}_{1z}+\hat{l}_{2z}$ is also conserved. This can be seen as follows. The two-body potential $V\left(r_{12}\right)$ can be decomposed as
\begin{equation}
V\left(r_{12}\right)=\sum_{l=-\infty}^{\infty}V_{l}\left(r_{1},r_{2}\right)\mathrm{exp}{[-il\left(\varphi_{1}-\varphi_{2}\right)]},
\end{equation}
with
\begin{equation}
V_{l}\left(r_{1},r_{2}\right)=\int rdrV\left(r\right)\int kdkJ_{0}\left(kr\right)J_{l}\left(kr_{1}\right)J_{l}\left(kr_{2}\right),
\end{equation}
in terms of the Bessel functions of the first kind $J_{n}(x)$ (see the details in Appendix~\ref{sec:AppA}). Then we have the commutation relation between $\hat{l}_{iz}$ and
$V\left(r_{12}\right)$ as
\begin{equation}
\left[\hat{l}_{iz},V\left(r_{12}\right)\right]=-i\hbar\frac{\partial V}{\partial r_{12}}\frac{\partial r_{12}}{\partial\varphi_{i}},
\end{equation}
which leads to
\begin{equation}
\left[\hat{L}_{z},V\left(r_{12}\right)\right]=0.
\end{equation}
As a consequence, the angular part of the two-body wave function is alternatively described by another four quantum numbers $\left\{ l_{z}\equiv\left(l_{1z}-l_{2z}\right)/2, S; L_{z}, S_{z}\right\} $, in which $L_{z}$ and $S_{z}$ are conserved. In the following, we are going to solve the two-body problem in the subspace of $L_{z}=0$ and $S_{z}=0$, and it gives $l_{1z}=-l_{2z}\equiv l$. The two-body wave function may be written as
\begin{equation}
\Psi\left({\bf r}_{1},{\bf r}_{2}\right)=\psi_{\uparrow\downarrow}\left({\bf r}_{1},{\bf r}_{2}\right)\left|\uparrow\downarrow\right\rangle +\psi_{\downarrow\uparrow}\left({\bf r}_{1},{\bf r}_{2}\right)\left|\downarrow\uparrow\right\rangle .\label{eq:2bWaveFunction}
\end{equation}
By inserting the two-body wave function Eq.~\eqref{eq:2bWaveFunction} into the Schr\"{o}dinger equation, we obtain
\begin{equation} \label{eq:2bodySE}
\begin{bmatrix}
\hat{H}_{\uparrow\downarrow}^{(0)}+V/2 & -V/2\\
-V/2 & \hat{H}_{\downarrow\uparrow}^{(0)}+V/2
\end{bmatrix}
\begin{bmatrix}
\psi_{\uparrow\downarrow}\\
\psi_{\downarrow\uparrow}
\end{bmatrix}=E
\begin{bmatrix}
\psi_{\uparrow\downarrow}\\
\psi_{\downarrow\uparrow}
\end{bmatrix}.
\end{equation}
\begin{widetext}
Regarding the spatial wave functions, we may expand them in the noninteracting basis as
\begin{equation} \label{eq:psi_ud}
\psi_{\uparrow\downarrow}\left({\bf r}_{1},{\bf r}_{2}\right)=
\sum_{k_{1}k_{2}l}A^{k_{1}k_{2}}_{l}u^{k_{1}}_{l,\uparrow}\left(r_{1}\right)u^{k_{2}}_{-l,\downarrow}\left(r_{2}\right)\frac{e^{il\phi}}{\sqrt{2\pi}},
\end{equation}
and
\begin{equation}
\psi_{\downarrow\uparrow}\left({\bf r}_{1},{\bf r}_{2}\right)=
\sum_{k_{1}k_{2}l}B^{k_{1}k_{2}}_{l}u^{k_{1}}_{l,\downarrow}\left(r_{1}\right)u^{k_{2}}_{-l,\uparrow}\left(r_{2}\right)\frac{e^{il\phi}}{\sqrt{2\pi}}.
\end{equation}
Recall that $l_{1z}=-l_{2z}\equiv l$ at the given total angular momentum $L_{z}=0$. Here, we have defined $\phi\equiv\varphi_{1}-\varphi_{2}$. After substituting these wave functions back into Eq.~\eqref{eq:2bodySE}, we obtain the following secular equation:
\begin{equation} \label{eq:2bSecularEq}
\left\{ 2\hbar\omega
\begin{bmatrix}
\left(k_{1}+k_{2}+\left|l-n\right|+1\right)\otimes\boldsymbol{\mathcal{I}} & 0\\
0 &\left(k_{1}+k_{2}+\left|l+n\right|+1\right)\otimes\boldsymbol{\mathcal{I}}
\end{bmatrix}
+
\begin{bmatrix}
\boldsymbol{\mathcal{V}}^{\left(\uparrow\downarrow\right)\left(\uparrow\downarrow\right)} & -\boldsymbol{\mathcal{V}}^{\left(\uparrow\downarrow\right)\left(\downarrow\uparrow\right)}\\
-\boldsymbol{\mathcal{V}}^{\left(\downarrow\uparrow\right)\left(\uparrow\downarrow\right)} & \boldsymbol{\mathcal{V}}^{\left(\downarrow\uparrow\right)\left(\downarrow\uparrow\right)}
\end{bmatrix}\right\} 
\begin{bmatrix}
\boldsymbol{\mathcal{A}}\\
\boldsymbol{\mathcal{B}}
\end{bmatrix}=E
\begin{bmatrix}
\boldsymbol{\mathcal{A}}\\
\boldsymbol{\mathcal{B}}
\end{bmatrix},
\end{equation}
\end{widetext}
with the identity matrix $\boldsymbol{\mathcal{I}}$ and the matrix $\boldsymbol{\mathcal{V}}$ (see the explicit form in Appendix~\ref{app:V-matrix}). It is easy to verify the relation $\left[\boldsymbol{\mathcal{V}}^{\left(\uparrow\downarrow\right)\left(\downarrow\uparrow\right)}\right]^{\dagger}=\boldsymbol{\mathcal{V}}^{\left(\downarrow\uparrow\right)\left(\uparrow\downarrow\right)}$. The concerned two-body spectrum $E$ as well as the wave functions are then obtained by numerically solving Eq.~\eqref{eq:2bSecularEq}.

\section{Numerical results} \label{sec: results}

For the convenience of numerical calculations, we choose a spherical-square-well (SSW) potential as the interaction potential, i.e.,
\begin{equation}
V\left(r_{12}\right)=\begin{cases}
-V_{0}, & 0\le r_{12}\le\epsilon, \\
0, & r_{12}>\epsilon,
\end{cases}
\end{equation}
with the depth $V_{0}>0$ and an interaction range $\epsilon$. The advantage of this choice lies in avoiding the complex regularization of the zero-range model and capturing the low-energy behavior of two-body states outside the interaction range, which should be universal for cold atoms.

\subsection{Without SOAM coupling}

As a self-examination, let us first consider the trivial system in the absence of SOAM coupling. It recovers exactly the case of two harmonically trapped atoms as described in Refs.~\cite{busch1998two,liu2010exact,peng2011high,peng2011non}. Here, we consider the spin degrees of freedom. This leads to additional degeneracy of energy levels in the two-body spectrum, which may be partially lifted when the SOAM coupling is present later. The c.m. motion is decoupled from the relative motion for two interacting atoms in a harmonic trap, and the combination of the c.m. energy $E_\mathrm{cm}$ and the relative-motion energy $E_\mathrm{rel}$ contributes to the total energy $E=E_\mathrm{cm}+E_\mathrm{rel}$. Explicitly, the c.m. motion energy takes the simple form of a harmonic oscillator as $E_\mathrm{cm}=\left(2n_{c}+\left|l_{c}\right|+1\right)\hbar\omega$, while the relative-motion energy is governed by the interaction in the spin-singlet channel via Eq.~\eqref{eq: Interaction} and is determined by the Schr\"{o}dinger equation:
\begin{equation}
\left[-\frac{\hbar^{2}}{2\mu}\nabla_{{\bf r}}^{2}+\frac{1}{2}\mu\omega^{2}r^{2}+V\left(r\right)\right]\psi\left({\bf r}\right)=E_\mathrm{rel}\psi\left({\bf r}\right).
\end{equation}
Note that the two-body interaction works only in the spin-singlet channel and does not affect the states in the spin-triplet channel. Here, $\mu=m/2$ is the reduced mass, and ${\bf{r}}\equiv {\bf{r}}_{12}$ is the relative coordinate of two atoms. Since the OAM $l_{r}$ of the relative motion is a good quantum number, different angular partial waves are decoupled. Thus, the wave function of the relative motion for the $l_{r}$th partial wave may be written as $\psi_{l_{r}}\left({\bf r}\right)=u_{_{l_{r}}}(r)\mathrm{exp}{(il_{r}\varphi)}/\sqrt{2\pi}$. After substituting back the wave function, the radial equation becomes
\begin{equation}
\left[\hat{H}_{r}^{(l_{r})}+V\left(r\right)\right]u_{_{l_{r}}}(r)=E_\mathrm{rel}u_{_{l_{r}}}(r),\label{eq: RelativeEq_NoSOC}
\end{equation}
where $\hat{H}_{r}^{(l_{r})}$ is the radial Hamiltonian of a 2D harmonic oscillator, i.e., 
\begin{equation}
\hat{H}_{r}^{(l_{r})}=-\frac{\hbar^{2}}{2\mu r}\frac{d}{dr}r\frac{d}{dr}+\frac{\hbar^{2}l_{r}^{2}}{2\mu r^{2}}+\frac{1}{2}\mu\omega^{2}r^{2}.
\end{equation}
\begin{figure}[t]
\includegraphics[width=0.48\textwidth]{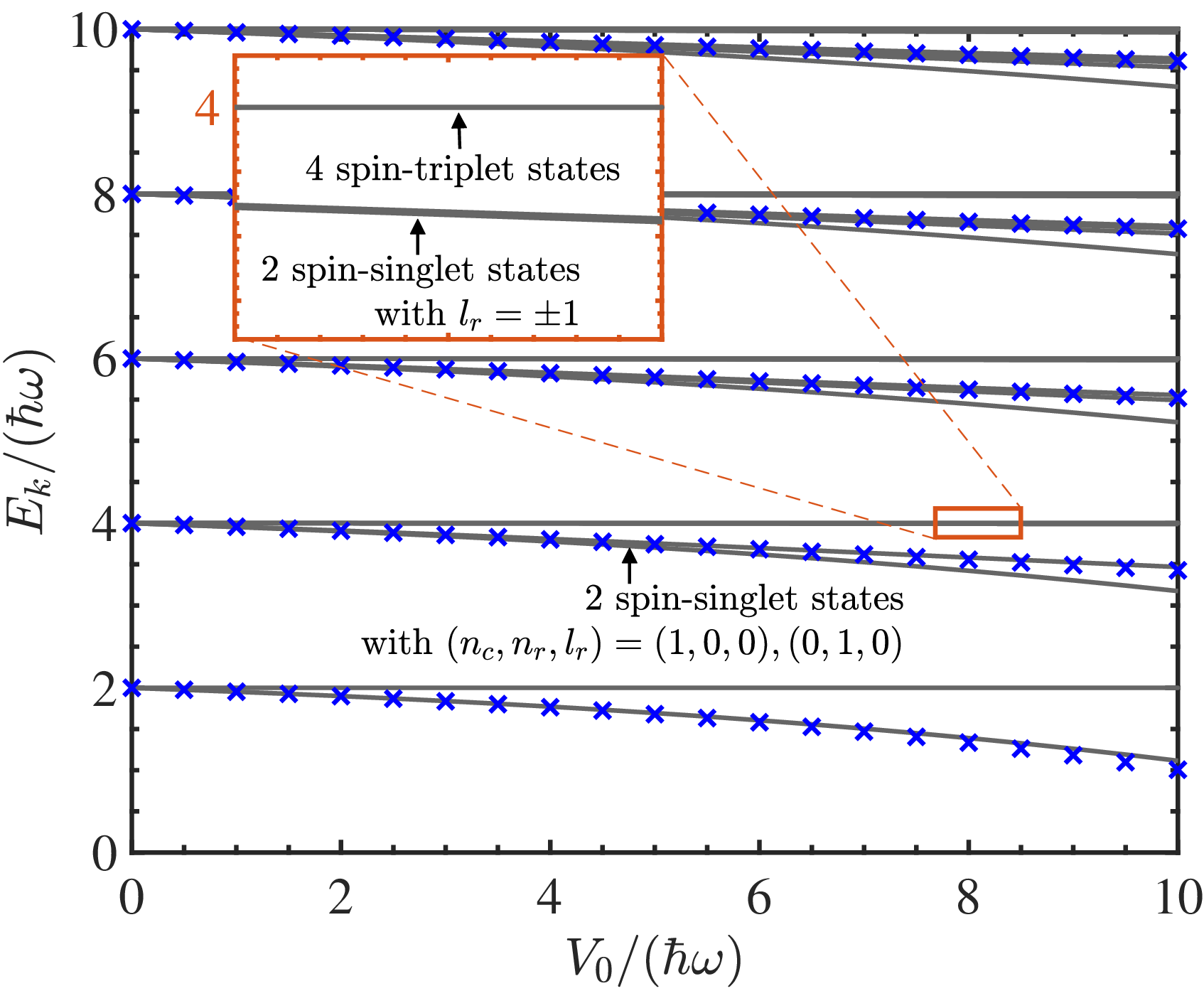}
\caption{(Color online) The energy spectrum of two atoms $E_k$ in the absence of spin-orbital-angular-momentum (SOAM) coupling as a function of the interaction strength denoted by the depth $V_0$ of a spherical-square-well potential, calculated by numerically solving Eq.~\eqref{eq:2bSecularEq} with $n=0$. The blue crosses denote the {\it s}-wave energy spectrum $E=E_\mathrm{rel}^{(s)}+\hbar \omega$ consisting of the ground center-of-mass (c.m.) energy with $n_c=0$ and the relative energy $E_\mathrm{rel}^{(s)}$ obtained by solving Eq.~\eqref{eq: EnergyEqRelativeMotion}. The inset zooms in the tiny region near $V_0=8\hbar\omega$ and $E_k=4\hbar\omega$ to emphasize four spin-triplet and two spin-singlet states. Here, we have set a relatively small interaction range $\epsilon=0.3a_\mathrm{ho}$. }
\label{figure2_En_V0}
\end{figure}
For self-consistency, we focus on the solution in the subspace of zero total OAM, i.e., $l_{c}=-l_{r}$. In the noninteracting limit, the energy of two atoms  takes the simple form of
\begin{equation} \label{eq: E_nc_nr}
E=2\left(n_{c}+n_{r}+\left|l_{r}\right|+1\right)\hbar\omega\equiv2\left(N+1\right)\hbar\omega,
\end{equation}
the degeneracy of which is $D_{N}=2\left(N+1\right)^2$. Taking the $N=1$ level with $E=4\hbar\omega$ as an example, there are eight degenerate states in total, four each in the spin-singlet and spin-triplet channels. Explicitly, these states, respectively, correspond to the quantum numbers $\left(n_c,n_r,l_r\right)=\left(1,0,0\right),\left(0,1,0\right),\left(0,0,\pm1\right)$. As the interaction is turned on, only the degeneracies for different energy levels in the spin-singlet channel are partially lifted, leaving the degeneracy of $l_r$ (for example, the degeneracy corresponding to $l_r=\pm1$ for the $N=1$ energy level remains, as shown in Fig.~\ref{figure2_En_V0}).

The radial equation of the relative motion in Eq.~\eqref{eq: RelativeEq_NoSOC} can be solved analytically for an SSW potential. Let us consider the ${\it s}$-wave solution ($l_{r}=0$) as an example. The solution outside the range of the interaction, i.e., $r>\epsilon$, takes the form of (unnormalized)
\begin{equation}
u_{0}^{>}\left(r\right)=\mathrm{exp}{\left(-\frac{r^{2}}{2d^{2}}\right)}U\left(-\nu,1,\frac{r^{2}}{d^{2}}\right),
\end{equation}
where $d=\sqrt{\hbar/\mu\omega}=\sqrt{2}a_\mathrm{ho}$ is the harmonic length for the relative motion, and $\nu$ constructs the ${\it s}$-wave relative-motion energy as
\begin{equation}
   E_\mathrm{rel}^{(s)}=\left(2\nu+1\right)\hbar\omega. 
\end{equation}
Here, $U\left(a,b,z\right)$ is the Kummer function of the second kind, which satisfies the boundary condition at a large distance, i.e., $u_{0}^{>}\left(r\to\infty\right)\sim0$. Inside the interaction range, the radial wave function has the form of
\begin{equation}
u_{0}^{<}\left(r\right)=c~\mathrm{exp}{\left(-\frac{r^{2}}{2d^{2}}\right)}M\left(-\kappa,1,\frac{r^{2}}{d^{2}}\right)
\end{equation}
with $\kappa=\nu+V_{0}/2\hbar\omega$. Here, $M\left(a,b,z\right)$ is the Kummer function of the first kind, which guarantees the wave function being finite at $r=0$, i.e., $u_{0}^{<}\left(r\to0\right)\sim\mathrm{const}\neq0$. By using the continuity condition of the radial wave function and its first-order derivative at $r=\epsilon$, we obtain the equation satisfied by the energy $E_\mathrm{rel}^{(s)}$
\begin{equation} \label{eq: EnergyEqRelativeMotion}
\nu\frac{U\left(1-\nu,2,\epsilon^{2}/2a_\mathrm{ho}^{2}\right)}{U\left(-\nu,1,\epsilon^{2}/2a_\mathrm{ho}^{2}\right)}+\kappa\frac{M\left(1-\kappa,2,\epsilon^{2}/2a_\mathrm{ho}^{2}\right)}{M\left(-\kappa,1,\epsilon^{2}/2a_\mathrm{ho}^{2}\right)}=0.
\end{equation}
After solving Eq.~\eqref{eq: EnergyEqRelativeMotion}, we could depict the total energy spectrum via $E=E_\mathrm{cm}+E_\mathrm{rel}^{(s)}$ for the $s$-partial wave in the absence of SOAM coupling.

\begin{figure*}[t]
\centering
\includegraphics[width=0.48\textwidth]{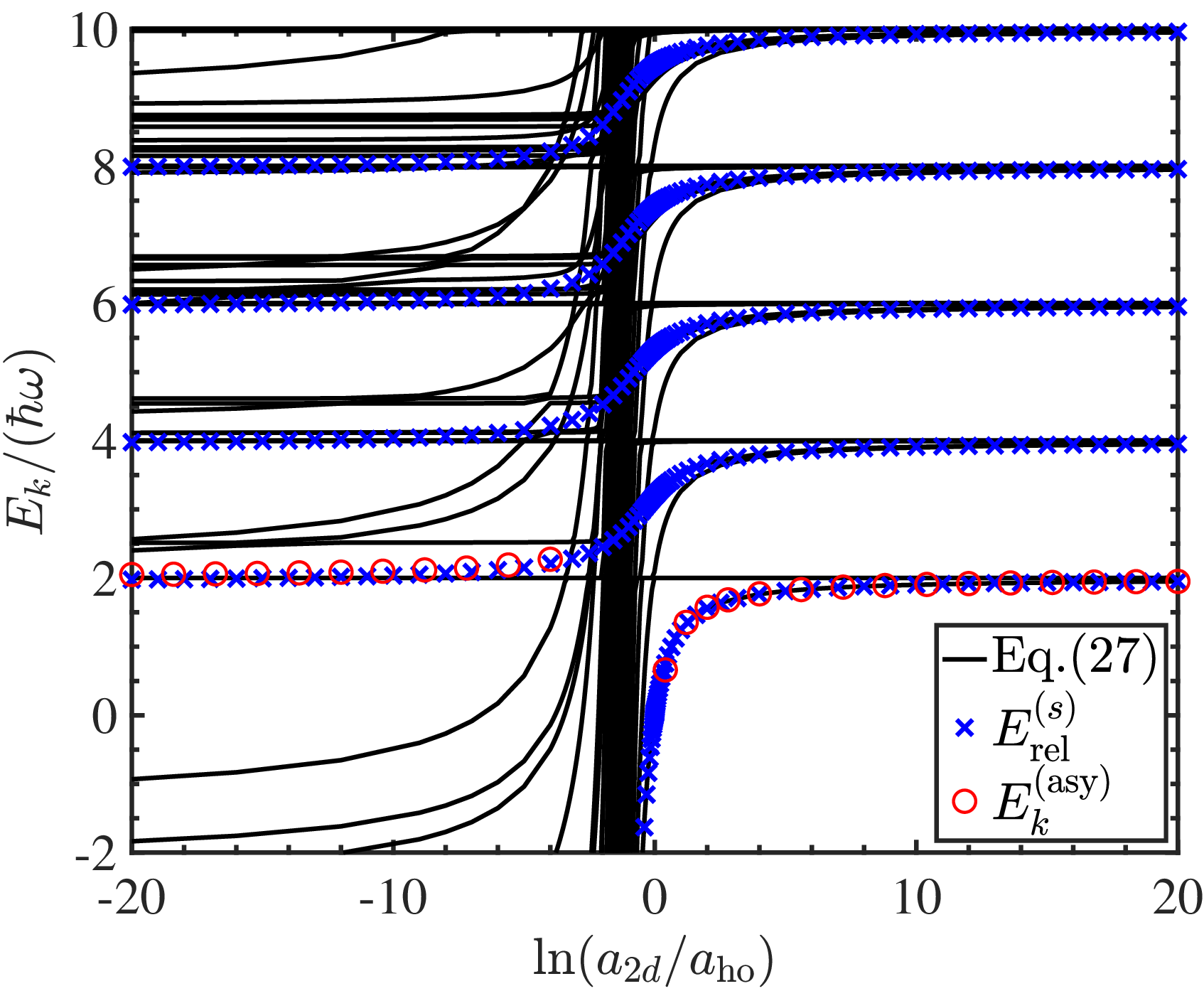}
\includegraphics[width=0.48\textwidth]{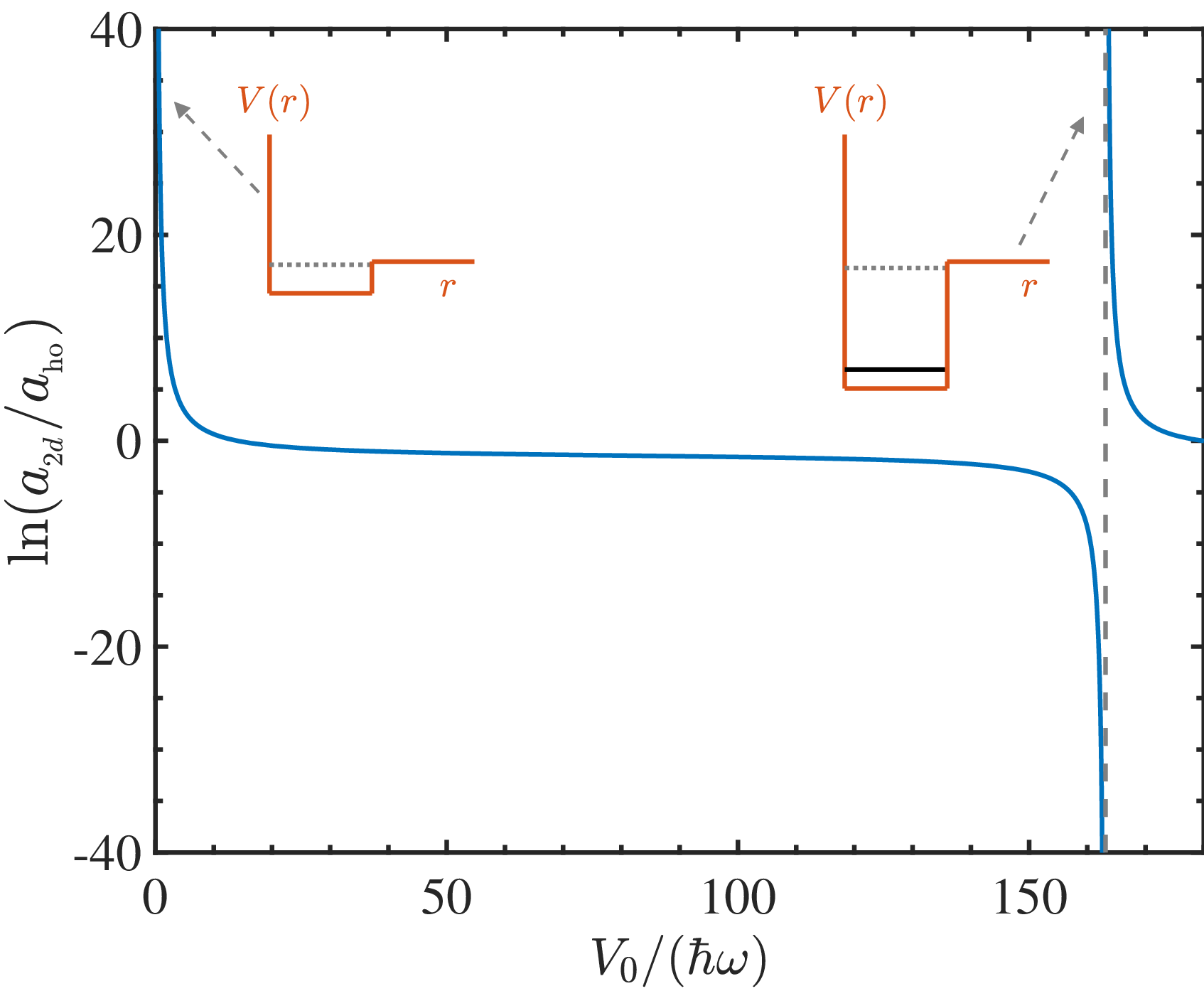}
\caption{(Color online) (Left) The energy spectrum of two atoms $E_k$ in the absence of spin-orbital-angular-momentum (SOAM) coupling as a function of the introduced 2D scattering
length $\ln{(a_{2d}/a_\mathrm{ho})}$ via Eq.~\eqref{eq:lna2d} for a spherical-square-well (SSW) potential. The solid lines are obtained by numerically solving Eq.~\eqref{eq:2bSecularEq} with $n=0$; the blue crosses denote the specific total energy $E_\mathrm{rel}^{(s)}+\hbar\omega$, with the {\it s}-wave relative energy $E_\mathrm{rel}^{(s)}$ calculated by solving the analytic expression Eq.~\eqref{eq: EnergyEqRelativeMotion}, while the red circles indicate the asymptotic behavior of the total energy $E^\mathrm{(asy)}_{k}$, i.e., Eq.~\eqref{eq: E_asymp}, in the noninteracting limit. (Right) The {\it s}-wave scattering length in 2D as a function of the SSW depth. The insets illustrate the position of resonance when a bound state appears near the threshold.}
\label{figure3_En_lna2dB_n0}
\end{figure*}

In Fig.~\ref{figure2_En_V0}, we present the typical two-body energy spectrum as a function of the SSW depth $V_0$ in a harmonic trap. The gray curves indicate the two-body spectrum obtained by numerically solving Eq.~\eqref{eq:2bSecularEq} in the absence of SOAM coupling, i.e., setting $n=0$. The blue crosses denote the {\it s}-wave energy spectrum $E=E_\mathrm{rel}^{(s)}+\hbar \omega$ consisting of the ground c.m. energy with $n_c=0$ and the relative energy $E_\mathrm{rel}^{(s)}$ obtained by solving Eq.~\eqref{eq: EnergyEqRelativeMotion}. As we anticipate, the degeneracies of energy levels in the spin-singlet channel are partially lifted by interaction while the spin-triplet states remain still. Taking the energy levels near $E=4\hbar\omega$ for example, there are four degenerate states in the spin-triplet channel, which are not affected by interaction. Instead, four previously degenerate states in the spin-singlet channel split into three as the SSW depth $V_0$ increases, two of which, corresponding to $l_r=\pm1$, are still degenerate, see the inset. In the figure, one can see that these analytical energies denoted by the blue crosses are in excellent agreement with the corresponding numerical results via Eq.~\eqref{eq:2bSecularEq} (i.e., lines counting for the $s$-wave case).

For ultracold atoms, the {\it s}-wave interaction is usually parameterized by the universal scattering length. Following Ref.~\cite{busch1998two}, we introduce a 2D scattering length for the SSW interaction by
\begin{equation} \label{eq:lna2d}
\ln\frac{a_{2d}}{a_\mathrm{ho}}=\frac{J_{0}\left(\sqrt{m\epsilon^{2}V_{0}/\hbar^{2}}\right)}{\sqrt{m\epsilon^{2}V_{0}/\hbar^{2}}J_{1}\left(\sqrt{m\epsilon^{2}V_{0}/\hbar^{2}}\right)}-\ln\frac{2a_\mathrm{ho}}{\epsilon e^{\gamma_{E}}},
\end{equation}
where $\gamma_E$ is the Euler gamma constant, and $J_{\nu}\left(\cdot\right)$ is the Bessel function of the first kind, see the detailed derivation in Appendix~\ref{app:scattering_length}. In Fig.~\ref{figure3_En_lna2dB_n0}, we present the two-body energy spectrum $E_k$ as a function of this introduced scattering length $\ln(a_{2d}/a_\mathrm{ho})$. Similarly, the solid curves indicate the total energy including different partial waves obtained by numerically solving Eq.~\eqref{eq:2bSecularEq}, while the blue crosses denote the analytically derived $s$-wave energy spectrum $E_\mathrm{rel}^{(s)}+\hbar\omega$ counting the ground c.m. energy. Furthermore, the {\it asymptotic} behavior of the $s$-wave energy spectrum $E^\mathrm{(asy)}_{k}$ in noninteracting limit around $E_k=2(k+1)\hbar\omega$ ($k=0,1,2,\cdots$) takes an explicit form of~\cite{busch1998two}
\begin{equation} \label{eq: E_asymp}
    E^\mathrm{(asy)}_{k} \cong 2\left(k+1\right)-\frac{2}{\ln{2}+2\ln{(a_{2d}/a_\mathrm{ho})}}\;,
\end{equation}
as denoted by red circles near the energy level $E_k=2\hbar\omega$ in the figure. Here, the ground-state energy of the c.m. motion is also included. In general, both the analytic results $E_\mathrm{rel}^{(s)}$ and $E^\mathrm{(asy)}_{k}$ show an excellent agreement with the corresponding ones in our numerical calculations. Unlike the situation in three dimensions, the two-body bound state appears even for an extremely shallow depth $V_0$ in 2D. This implies a positive 2D scattering length for an arbitrary SSW depth. The scattering resonance occurs, corresponding to $\ln \left(a_{2d}/a_{\text{ho}}\right)\to+\infty$ or $V_0\to0$ as shown in the right plot of Fig.~\ref{figure3_En_lna2dB_n0}, in the noninteracting limit, once the bound state appears. Thus, the energy spectrum tends to the noninteracting result of Eq.~\eqref{eq: E_nc_nr}. The binding energy of the two-body bound state increases (but negative) as the SSW depth increases. During this process, the 2D scattering length decreases to zero, i.e., $\ln \left(a_{2d}/a_{\text{ho}}\right)\to-\infty$, before the next two-body bound state appears. The energy spectrum then tends to approach the noninteracting result again.

\subsection{With SOAM coupling}

We now turn to discuss the role of SOAM coupling by considering a nonzero $n$ (the OAM transferred to atoms in the Raman process) and solving Eq.~\eqref{eq:2bSecularEq} numerically. 

\subsubsection{Energy spectrum} \label{sec:EnergySpectrum}

\begin{figure}[t]
\centering
\includegraphics[width=0.48\textwidth]{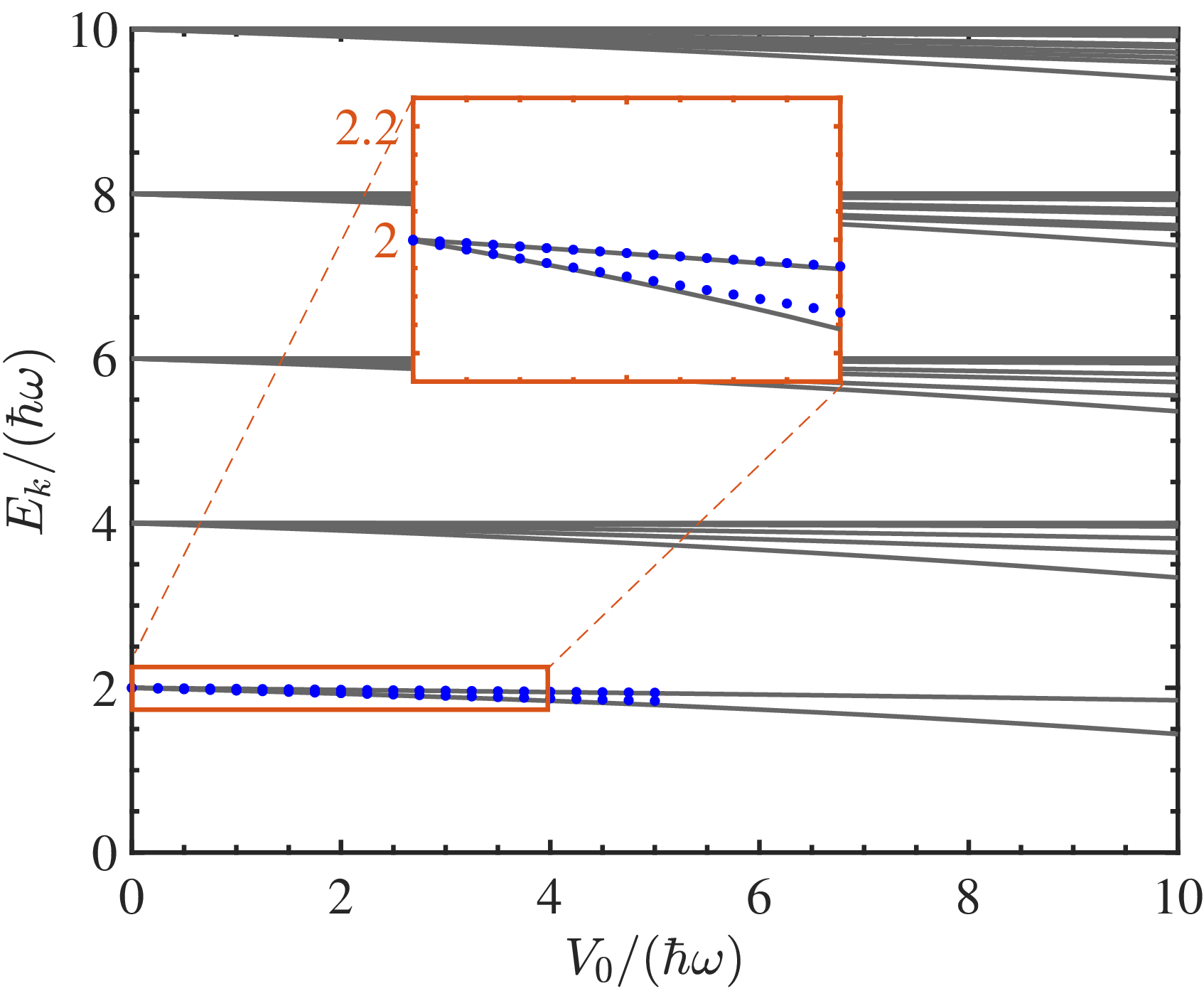}
\caption{(Color online) The energy spectrum of two atoms $E_k$ in the presence of spin-orbital-angular-momentum (SOAM) coupling as a function of the spherical-square-well depth $V_{0}$, calculated by numerically solving Eq.~\eqref{eq:2bSecularEq} with $n=2$. The blue dots indicate the lowest two energy levels at small $V_0$, calculated from the perturbation approach. The inset zooms in the tiny region near $E_k=2\hbar\omega$ with $V_0/\hbar\omega$ varying from 0 to 4.}
\label{figure4_En_V0_soamc_n2}
\end{figure}

After considering the SOAM coupling by setting $n=2$, the calculated two-body energy spectrum is presented as a function of the SSW depth $V_{0}$ in Fig.~\ref{figure4_En_V0_soamc_n2}. Under the interplay between the SOAM coupling and interaction, the degeneracy of each energy level is further lifted. This can be understood as follows. The interaction lifts the degeneracy of the energy levels in the spin-singlet channel, while the energy spectrum in the spin-triplet channel is unchanged, as we have found in the absence of SOAM coupling. However, the SOAM coupling will mix the spin-singlet and spin-triplet states since the total spin is no longer conserved as expected. As a consequence, it gives rise to the additional splitting of energy levels and the elimination of energy degeneracies. 

To further understand the underlying physics of the two-body spectrum in the presence of SOAM coupling, let us adopt a perturbation analysis of Eq.~\eqref{eq: Hamiltonian_sub} in the weakly interacting limit, i.e., $V_{0}/\hbar\omega\ll 1$. In the absence of interaction, i.e., $V_0=0$, the Hamiltonian $\hat{H}$ is readily diagonalized in the noninteracting basis, i.e., $\left\{ \left|\psi_{\uparrow\downarrow}\right\rangle \left|\uparrow\downarrow\right\rangle ,\left|\psi_{\uparrow\downarrow}\right\rangle \left|\uparrow\downarrow\right\rangle \right\}$,
where the spatial part takes the form of
\begin{subequations}
\begin{eqnarray}
\psi_{\uparrow\downarrow}\left({\bf r}_{1},{\bf r}_{2}\right) & = & u_{l,\uparrow}^{k_{1}}\left(r_{1}\right)u_{-l,\downarrow}^{k_{2}}\left(r_{2}\right)\frac{e^{il\phi}}{2\pi},
\\
\psi_{\downarrow\uparrow}\left({\bf r}_{1},{\bf r}_{2}\right) & = & u_{l,\downarrow}^{k_{1}}\left(r_{1}\right)u_{-l,\uparrow}^{k_{2}}\left(r_{2}\right)\frac{e^{il\phi}}{2\pi},
\end{eqnarray}
\end{subequations}
which can be denoted as $\left|\psi_{\uparrow\downarrow\left(\downarrow\uparrow\right)}\right\rangle \equiv\left|k_{1},k_{2};l\right\rangle $.
Here, we note that $l_{1}=-l_{2}\equiv l$ is required by the zero
total OAM and $\phi=\phi_{1}-\phi_{2}$. Specifically,
we have
\begin{equation}
\begin{bmatrix}
\left\langle \psi_{\uparrow\downarrow}\right|, & \left\langle \psi_{\downarrow\uparrow}\right|
\end{bmatrix}
\begin{bmatrix}
\hat{H}_{\uparrow\downarrow}^{(0)} & 0\\
0 & \hat{H}_{\downarrow\uparrow}^{(0)}
\end{bmatrix}
\begin{bmatrix}
\left|\psi_{\uparrow\downarrow}\right\rangle \\
\left|\psi_{\uparrow\downarrow}\right\rangle 
\end{bmatrix} = 
\begin{bmatrix}
E_{\uparrow\downarrow}^{(0)} & 0\\
0 & E_{\downarrow\uparrow}^{(0)}
\end{bmatrix}
\end{equation}
in the spin basis $\left\{ \left|\uparrow\downarrow\right\rangle ,\left|\uparrow\downarrow\right\rangle \right\}$,
with
\begin{eqnarray}
E_{\uparrow\downarrow}^{(0)} & = & 2\left(k_{1}+k_{2}+\left|l-n\right|+1\right)\hbar\omega,\\
E_{\downarrow\uparrow}^{(0)} & = & 2\left(k_{1}+k_{2}+\left|l+n\right|+1\right)\hbar\omega.
\end{eqnarray}
It is straightforward to see that the energy spectrum
becomes even times of the harmonic trapping energy as $E^{(0)}=2\hbar\omega,4\hbar\omega,...$.
Let us focus the discussion on the lowest two energy levels as an
example, i.e., degenerate $E_{\uparrow\downarrow}^{(0)}=E_{\downarrow\uparrow}^{(0)}=2\hbar\omega$,
corresponding to the states $\left|\psi_{\uparrow\downarrow}\right\rangle =\left|0,0;n\right\rangle$
and $\left|\psi_{\downarrow\uparrow}\right\rangle =\left|0,0;-n\right\rangle$.
Therefore, we have the lowest energy $E_{0}^{(0)}=2\hbar\omega$ with
a twofold degeneracy, manifested as
\begin{subequations}
\begin{eqnarray}
\hat{H}^{(0)}\left|\Psi_{1}^{(0)}\right\rangle  & = & E_{0}^{(0)}\left|\Psi_{1}^{(0)}\right\rangle,
\\
\hat{H}^{(0)}\left|\Psi_{2}^{(0)}\right\rangle  & = & E_{0}^{(0)}\left|\Psi_{2}^{(0)}\right\rangle, 
\end{eqnarray}
\end{subequations}
with
\begin{subequations}\label{zero-order WF}
\begin{eqnarray}
\Psi_{1}^{(0)}\left({\bf r}_{1},{\bf r}_{2}\right) & = & \frac{e^{-\left(r_{1}^{2}+r_{2}^{2}\right)/2a_\mathrm{ho}^{2}}}{\pi a_\mathrm{ho}^{2}}e^{+in\varphi}\left|\uparrow\downarrow\right\rangle ,\\
\Psi_{2}^{(0)}\left({\bf r}_{1},{\bf r}_{2}\right) & = & \frac{e^{-\left(r_{1}^{2}+r_{2}^{2}\right)/2a_\mathrm{ho}^{2}}}{\pi a_\mathrm{ho}^{2}}e^{-in\varphi}\left|\uparrow\downarrow\right\rangle .
\end{eqnarray}
\end{subequations}

When the interaction $V$ is turned on but small, we may utilize the degenerate perturbation theory to see how this twofold degeneracy of the lowest energy is lifted. The unperturbed wave function $\left|\Psi^{(0)}\right\rangle $ could be expanded as a linear combination of $\left\{ \left|\Psi_{1}^{(0)}\right\rangle ,\left|\Psi_{2}^{(0)}\right\rangle \right\} $ in the degenerate sub-Hilbert space, i.e.,
\begin{equation}
\left|\Psi^{(0)}\right\rangle =\alpha_{1}\left|\Psi_{1}^{(0)}\right\rangle +\alpha_{2}\left|\Psi_{2}^{(0)}\right\rangle .
\label{psi0}
\end{equation}
The perturbed Hamiltonian takes the form of
\begin{equation}
\hat{H}=\hat{H}^{(0)}+\hat{H}^{(1)},
\end{equation}
with the perturbation
\begin{equation}
\hat{H}^{(1)}=\left[\begin{array}{cc}
+V/2 & -V/2\\
-V/2 & +V/2
\end{array}\right].
\end{equation}
Up to the first-order approximation, the perturbed wave function and the corresponding energy can be written as
\begin{subequations}
\begin{eqnarray}
\left|\Psi\right\rangle  & \approx & \left|\Psi^{(0)}\right\rangle +\left|\Psi^{(1)}\right\rangle , \\
E_{0} & \approx & E_{0}^{(0)}+E_{0}^{(1)}, \label{eq:E_perturb}
\end{eqnarray}
\end{subequations}
where $\left|\Psi^{(1)}\right\rangle $ and $E_{0}^{(1)}$ are the first-order corrections to the wave function and energy, respectively. After substituting the perturbed wave function and the corresponding energy back into the Schr\"{o}dinger equation $\hat{H}\left|\Psi\right\rangle =E_{0}\left|\Psi\right\rangle $, we get
\begin{equation}
\hat{H}^{(0)}\left|\Psi^{(1)}\right\rangle +\hat{H}^{(1)}\left|\Psi^{(0)}\right\rangle \approx E_{0}^{(0)}\left|\Psi^{(1)}\right\rangle +E_{0}^{(1)}\left|\Psi^{(0)}\right\rangle,
\end{equation}
up to the first-order approximation. After taking the inner product respectively with $\left\langle \Psi_{1}^{(0)}\right|$
and $\left\langle \Psi_{2}^{(0)}\right|$, we obtain eventually the secular equation:
\begin{equation}\label{Wmatrix}
\begin{bmatrix}
W_{11} & W_{12}\\
W_{21} & W_{22}
\end{bmatrix}
\begin{bmatrix}
\alpha_{1}\\
\alpha_{2} 
\end{bmatrix}
=E_{0}^{(1)}
\begin{bmatrix}
\alpha_{1}\\
\alpha_{2} 
\end{bmatrix},
\end{equation}
with the matrix element being
\begin{equation}
W_{mn}\equiv\left\langle \Psi_{m}^{(0)}\right|H^{(1)}\left|\Psi_{n}^{(0)}\right\rangle ,
\end{equation}
which is straightforwardly constructed by using Eq.~\eqref{zero-order WF}. Then we find that the matrix ${\bf W}$ takes the form of
\begin{equation}
{\bf W}=
\begin{bmatrix}
\mathcal{F} & \mathcal{G}\\
\mathcal{G} & \mathcal{F}
\end{bmatrix}V_{0},
\end{equation}
with the dimensionless constants
\begin{subequations}
\begin{eqnarray}
\mathcal{F} & = & \frac{-2\epsilon}{a_\mathrm{ho}^{4}}\int dkJ_{1}\left(k\epsilon\right)\left[\int r e^{-r^{2}/2a_\mathrm{ho}^{2}}J_{0}\left(kr\right)\right]^{2},\\
\mathcal{G} & = & \frac{+2\epsilon}{a_\mathrm{ho}^{4}}\int dk J_{1}\left(k\epsilon\right)\left[\int  r e^{-r^{2}/2a_\mathrm{ho}^{2}}J_{2n}\left(kr\right)\right]^{2},
\end{eqnarray}
\end{subequations}
where $\epsilon$ is the interaction range. The straightforward diagonalization
of ${\bf W}$ yields the first-order correction of the energy:
\begin{equation}
E_{\pm}^{(1)}=\left(\mathcal{F}\pm\mathcal{G}\right)V_{0},
\end{equation}
and the corresponding eigenvectors:
\begin{equation}
\begin{bmatrix}
\alpha_{1}^{(\pm)}\\
\alpha_{2}^{(\pm)}
\end{bmatrix}=
\begin{bmatrix}
1/\sqrt{2}\\
\pm1/\sqrt{2}
\end{bmatrix}.
\end{equation}
Therefore, the zero-order wave function in Eq.~\eqref{psi0} has the form of
\begin{equation}
\left|\Psi_{\pm}^{(0)}\right\rangle =\frac{1}{\sqrt{2}}\left(\left|\Psi_{1}^{(0)}\right\rangle \pm\left|\Psi_{2}^{(0)}\right\rangle \right).
\end{equation}

In Fig.~\ref{figure4_En_V0_soamc_n2} and the inset, the corrected energy $E_{0}\approx E_{0}^{(0)}+E_{0}^{(1)}$ of the $E_{0}^{(0)}=2\hbar\omega$ level is denoted by blue dots in the weakly interacting limit. We can find that the lowest two energy levels split as interaction strengthens and coincides with the exact numerical results shown in gray lines at small values of $V_0$.
In the absence of SOAM coupling, i.e., $n=0$, we have
\begin{equation}
\mathcal{G}=-\mathcal{F}=\frac{1-e^{-\epsilon^{2}/2a_\mathrm{ho}^{2}}}{2},
\end{equation}
and 
\begin{equation}
\psi_{\uparrow\downarrow}\left({\bf r}_{1},{\bf r}_{2}\right)=\psi_{\downarrow\uparrow}\left({\bf r}_{1},{\bf r}_{2}\right)=\frac{e^{-\left(r_{1}^{2}+r_{2}^{2}\right)/2a_\mathrm{ho}^{2}}}{\pi a_\mathrm{ho}^{2}}.
\label{V=0}
\end{equation}
In this case, the first-order correction of energy becomes
\begin{equation}
E_{+}=0,\quad E_{-}=-2\mathcal{G}V_{0}=\left(e^{-\epsilon^{2}/2a_\mathrm{ho}^{2}}-1\right)V_{0},
\end{equation}
and the corresponding zero-order wave function $\Psi^{(0)}\left({\bf r}_{1},{\bf r}_{2}\right)$
has the form of
\begin{equation}
\Psi_{\pm}^{(0)}\left({\bf r}_{1},{\bf r}_{2}\right)=\frac{e^{-\left(r_{1}^{2}+r_{2}^{2}\right)/2a_\mathrm{ho}^{2}}}{\pi a_\mathrm{ho}^{2}}\cdot\frac{1}{\sqrt{2}}\left(\left|\uparrow\downarrow\right\rangle \pm\left|\downarrow\uparrow\right\rangle \right),
\end{equation}
which is simply the wave function in the spin-singlet and spin-triplet channels. We can see that the energy in the spin-triplet channel is not affected by the interaction as we anticipated since the interaction only exists in the spin-singlet channel. However, when the SOAM coupling is present, the spin-singlet and spin-triplet states become mixed. Both energies $E_{\pm}$ are shifted by the interaction, as shown in Fig.~\ref{figure4_En_V0_soamc_n2}.
\begin{figure}[t]
\centering
\includegraphics[width=0.48\textwidth]{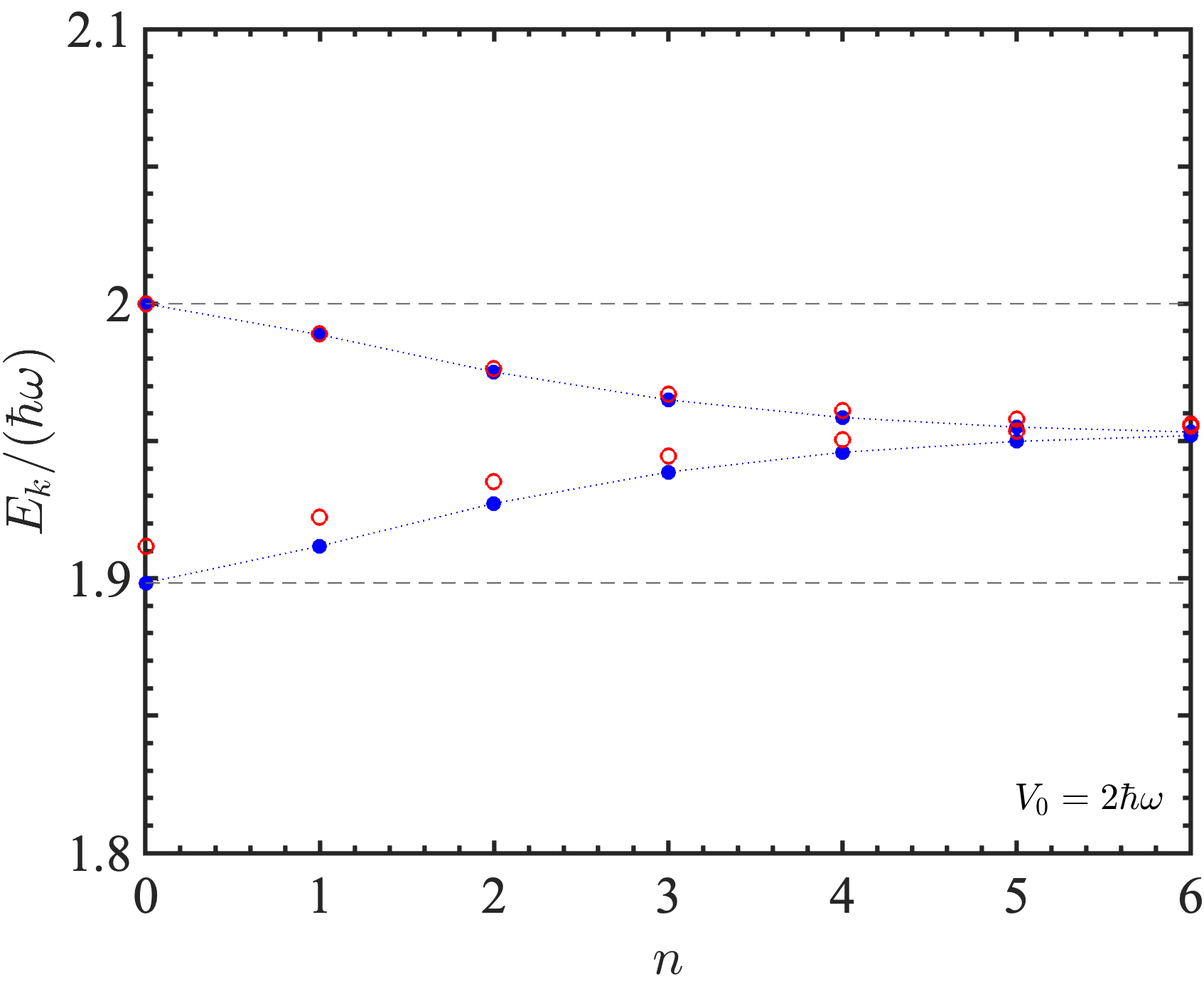}
\caption{(Color online) Lowest two energy levels $E_k$ of two atoms in a harmonic trap with the SOAM coupling, as a function of the transferred orbital angular momentum $n$ at a small spherical-square-well depth $V_0=2\hbar\omega$. The numerical results in blue dots are compared with the one calculated from the perturbation approach denoted by red circles.}
\label{figure5_spectrum_nr}
\end{figure}

By taking a relatively small interaction strength $V_{0}=2\hbar\omega$, we further illustrate these two lowest energy levels as functions of the transferred OAM $n$ in Fig.~\ref{figure5_spectrum_nr}. The numerically calculated energies in blue dots show a great agreement with the one from the perturbation approach denoted by red circles. Surprisingly, we find that the separate energy levels due to interactions tend to approach each other and restore approximately the degeneracy as $n$ increases. This can be explained by the perturbation theory as follows. To the first-order approximation, we find that the off-diagonal elements $W_{12}=W_{21}$ of the matrix $W$ in Eq.~\eqref{Wmatrix} decrease toward zero as $n$ increases, while the diagonal elements $W_{11}=W_{22}$ are irrelevant to $n$. Therefore, the off-diagonal elements are becoming negligible compared with the diagonal elements at large $n$, leading to no level repulsion and a tendency of degeneracy. The weak interaction introduces only a uniform shift to previously splitting energy levels.

\subsubsection{Correlations} \label{sec:Correlations}

\begin{figure}[t]
\includegraphics[width=0.48\textwidth]{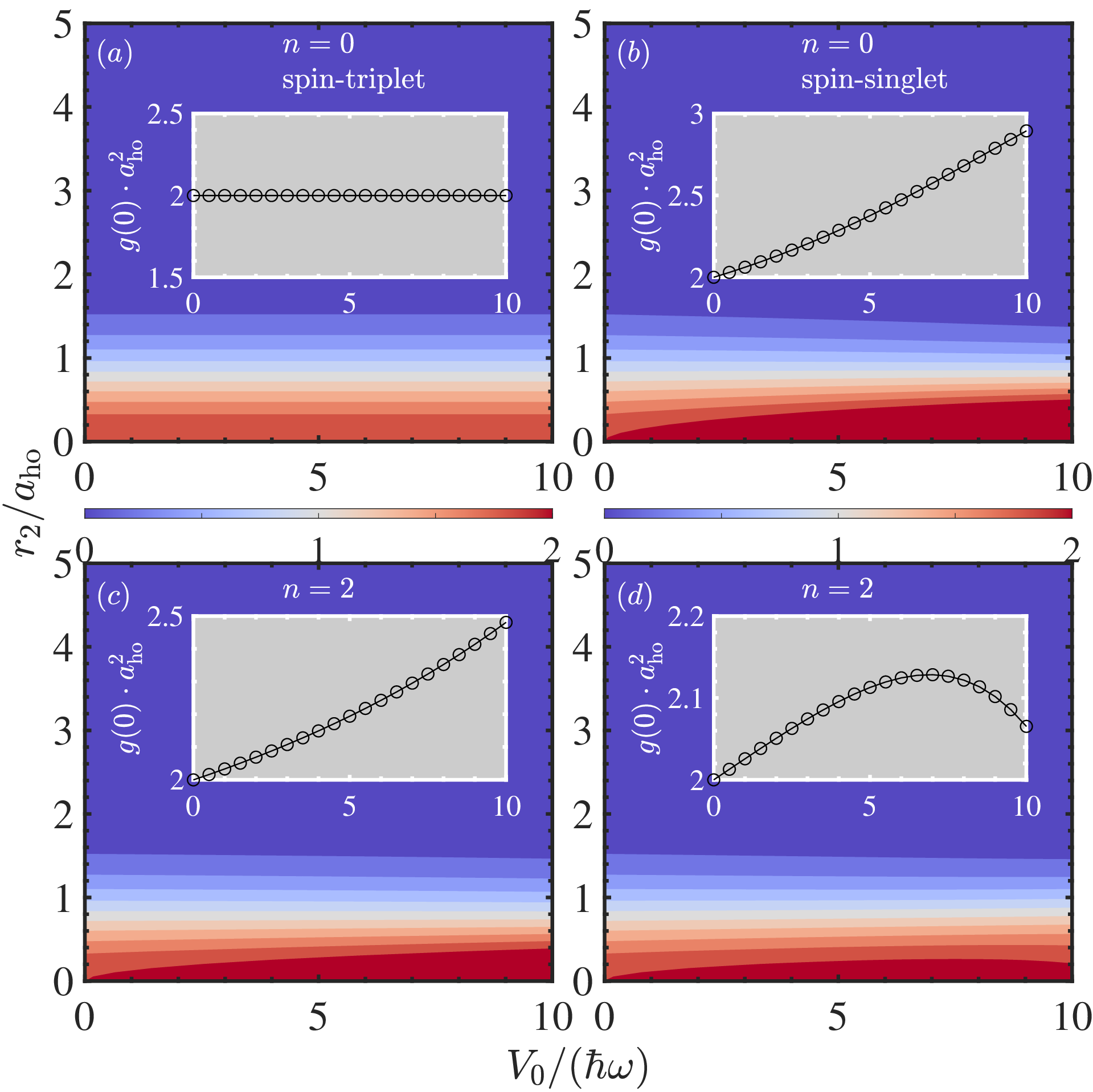}
\caption{(Color online) Contour plots of correlation functions $g\left(r_{2}\right)$ of two lowest states for energy level $E_k=2\hbar\omega$, i.e., second-lowest and lowest energy levels, respectively, in Figs.~\ref{figure2_En_V0} and~\ref{figure4_En_V0_soamc_n2}, as functions of radius $r_2$ and the spherical-square-well depth $V_0$. The color gradient varying from blue to red indicates the corresponding correlation function value increasing from 0. In the insets, the specific values of correlation
functions at $r_2=0$ are depicted as a function of $V_0$.}
\label{figure6_correlation}
\end{figure}

We turn to discuss the correlations of two atoms by introducing a correlation function or an integrated density function. If we fix the position of atom $1$, for example, at ${\bf r}_{1}$, the probability of finding atom $2$ at position ${\bf r}_{2}$ is $\left|\Psi\left({\bf r}_{1},{\bf r}_{2}\right)\right|^{2}$. Therefore, the total probability of atom $2$ appearing at position ${\bf r}_{2}$ is simply obtained by integrating over all ${\bf r}_{1}$, and that is
\begin{equation}
g\left({\bf{r}}_{2}\right)=\int d{\bf r}_{1}\left|\Psi\left({\bf r}_{1},{\bf r}_{2}\right)\right|^{2}.
\end{equation}
The function $g\left({\bf{r}}_{2}\right)$ characterizes the intrinsic correlation between the two atoms and has the dimension of length$^{-2}$. In Fig.~\ref{figure6_correlation}, the correlation function $g\left(r_{2}\right)$ is plotted as a function of $r_{2}=\left|{\bf{r}}_{2}\right|$ and the SSW depth $V_{0}$ for the two lowest-energy states around $E=2\hbar\omega$. A rising correlation function value from 0 is represented by the color gradient varying from blue to red. Here, we have also integrated out the angular part with respect to $\varphi_{2}$. 

In the absence of SOAM coupling, i.e., $n=0$ shown by the upper panel in Fig.~\ref{figure6_correlation}, the system in the noninteracting limit is nothing but two free atoms. Previously, we have obtained the zero-order approximation of the wave
function in Eq.~\eqref{V=0}. Then the correlation function in the noninteracting
limit simply takes the form of 
\begin{equation}
g^{(0)}\left({\bf r}_{2}\right)=\int d{\bf r}_{1}\left|\Psi_{\pm}^{(0)}\left({\bf r}_{1},{\bf r}_{2}\right)\right|^{2}=\frac{e^{-r_{2}^{2}/a_\mathrm{ho}^{2}}}{\pi a_\mathrm{ho}^{2}},
\end{equation}
and $g^{(0)}\left(r_{2}\right)=2\mathrm{exp}{(-r_{2}^{2}/a_\mathrm{ho}^{2})}/a_\mathrm{ho}^{2}$ by further integrating over $\varphi_{2}$, which yields $g\left(r_{2}=0\right)=2$
in both spin-singlet and spin-triplet channels. The correlation in the spin-singlet channel increases as the interaction strength increases, while it remains unchanged in the spin-triplet channel as anticipated since the interaction appears only in the spin-singlet channel, i.e., Eq.~\eqref{eq: Interaction}, which are clearly shown in the insets of Figs.~\ref{figure6_correlation}(a) and~\ref{figure6_correlation}(b). 
 
In sharp contrast, when the SOAM-coupling effect is included, e.g., $n=2$ shown by the lower panel in Fig.~\ref{figure6_correlation}, the states in the spin-singlet and spin-triplet channels are closely coupled. This can be seen from the zero-order perturbed wave function, i.e.,
\begin{equation}
\left|\Psi_{\pm}^{(0)}\right\rangle =\frac{e^{-\left(r_{1}^{2}+r_{2}^{2}\right)/2a_\mathrm{ho}^{2}}}{\sqrt{2}\pi a_\mathrm{ho}^{2}}\left(e^{+i2\varphi}\left|\uparrow\downarrow\right\rangle \pm e^{-i2\varphi}\left|\downarrow\uparrow\right\rangle \right),
\end{equation}
which is a mixture of those in the spin-singlet and spin-triplet channels,
while the correlation function is again $g^{(0)}\left(r_{2}\right)=2\mathrm{exp}{(-r_{2}^{2}/a_\mathrm{ho}^{2})}/a_\mathrm{ho}^{2}$
in the noninteracting limit. In the insets of Figs.~\ref{figure6_correlation}(c) and~\ref{figure6_correlation}(d), we find that the SOAM coupling plays a crucial role, and both correlation functions at $r_2=0$ of these two lowest-energy states exhibit a strong dependence on the interaction strength.

\section{Conclusions} \label{sec: summary}

We have studied the roles of SOAM coupling and a two-body interaction potential in the underlying physics of two harmonically trapped atoms by addressing the associated two-body energy spectrum and correlations. Starting from the two-body Hamiltonian counting a tunable interaction potential and the SOAM coupling, we have derived an explicit secular equation for numerically calculating the associated eigenenergy and eigenfunction. In the absence of SOAM coupling, the calculated energy spectrum can reproduce well the analytic result in previous works, i.e., only the spin-singlet states are significantly affected by the two-body interaction and the degeneracies in the energy spectrum disappear gradually as the interaction strength rises. 

In the presence of SOAM coupling, we have made a careful analysis of the two-body spectrum and the correlation as functions of the interaction strength as well as the transferred OAM. We first develop a perturbation approach to justify the numerical result in the weak-interaction limit, which demonstrates the crucial role of the interplay between the interaction and SOAM coupling in the elimination of energy-level degeneracy as the interaction enhances. In addition, we have introduced a correlation function, i.e., an integrated one-body density function, to characterize the behavior of two-body wave functions with respect to the interaction strength as well as the radius with and without SOAM coupling. The correlations show direct evidence of the mixing of the spin-triplet and spin-singlet wave functions due to SOAM coupling. At large transferred angular momentum, we find that the deviated energy levels of the lowest two states due to a definite weak interaction tend to approach each other and restore the degeneracy as the strength of SOAM coupling increases. This intriguing behavior is further explained by employing the perturbation approach. 

We have focused our study in the subspace of two definite conserved quantum numbers, corresponding to vanishing total OAM $L_{z}$ and total spin $S_{z}$ along the $z$ axis. This could be conveniently generalized to nonzero values of $L_{z}$ and $S_{z}$. In addition, the transverse effective Zeeman field has been neglected here due to the weak light-atom coupling in space in the present experimental setup, which may also be further considered according to the exact diagonalization method.   

\begin{acknowledgments}
This paper is supported by the Natural Science Foundation of China under Grant No. 12204413 and the Science Foundation of Zhejiang Sci-Tech University under Grant No. 21062339-Y (X.-L.C.), and the Natural Science Foundation of China under Grant No. 12374250, National Key R$\&$D Program under Grant No. 2022YFA1404102, and Innovation Program for Quantum Science and Technology under Grant No. 2023ZD0300401 (S.-G.P.).
\end{acknowledgments}

\bibliography{soamc2body}

\providecommand{\noopsort}[1]{}\providecommand{\singleletter}[1]{#1}%
\begin{thebibliography}{61}%
\makeatletter
\providecommand \@ifxundefined [1]{%
 \@ifx{#1\undefined}
}%
\providecommand \@ifnum [1]{%
 \ifnum #1\expandafter \@firstoftwo
 \else \expandafter \@secondoftwo
 \fi
}%
\providecommand \@ifx [1]{%
 \ifx #1\expandafter \@firstoftwo
 \else \expandafter \@secondoftwo
 \fi
}%
\providecommand \natexlab [1]{#1}%
\providecommand \enquote  [1]{``#1''}%
\providecommand \bibnamefont  [1]{#1}%
\providecommand \bibfnamefont [1]{#1}%
\providecommand \citenamefont [1]{#1}%
\providecommand \href@noop [0]{\@secondoftwo}%
\providecommand \href [0]{\begingroup \@sanitize@url \@href}%
\providecommand \@href[1]{\@@startlink{#1}\@@href}%
\providecommand \@@href[1]{\endgroup#1\@@endlink}%
\providecommand \@sanitize@url [0]{\catcode `\\12\catcode `\$12\catcode `\&12\catcode `\#12\catcode `\^12\catcode `\_12\catcode `\%12\relax}%
\providecommand \@@startlink[1]{}%
\providecommand \@@endlink[0]{}%
\providecommand \url  [0]{\begingroup\@sanitize@url \@url }%
\providecommand \@url [1]{\endgroup\@href {#1}{\urlprefix }}%
\providecommand \urlprefix  [0]{URL }%
\providecommand \Eprint [0]{\href }%
\providecommand \doibase [0]{https://doi.org/}%
\providecommand \selectlanguage [0]{\@gobble}%
\providecommand \bibinfo  [0]{\@secondoftwo}%
\providecommand \bibfield  [0]{\@secondoftwo}%
\providecommand \translation [1]{[#1]}%
\providecommand \BibitemOpen [0]{}%
\providecommand \bibitemStop [0]{}%
\providecommand \bibitemNoStop [0]{.\EOS\space}%
\providecommand \EOS [0]{\spacefactor3000\relax}%
\providecommand \BibitemShut  [1]{\csname bibitem#1\endcsname}%
\let\auto@bib@innerbib\@empty
\bibitem [{\citenamefont {Landau}\ and\ \citenamefont {Lifshitz}(2013)}]{Landau2007Q}%
  \BibitemOpen
  \bibfield  {author} {\bibinfo {author} {\bibfnamefont {L.~D.}\ \bibnamefont {Landau}}\ and\ \bibinfo {author} {\bibfnamefont {E.~M.}\ \bibnamefont {Lifshitz}},\ }\href@noop {} {\emph {\bibinfo {title} {Quantum mechanics: non-relativistic theory}}},\ Vol.~\bibinfo {volume} {3}\ (\bibinfo  {publisher} {Elsevier},\ \bibinfo {year} {2013})\BibitemShut {NoStop}%
\bibitem [{\citenamefont {De-Shalit}\ and\ \citenamefont {Talmi}(2013)}]{de2013nuclear}%
  \BibitemOpen
  \bibfield  {author} {\bibinfo {author} {\bibfnamefont {A.}~\bibnamefont {De-Shalit}}\ and\ \bibinfo {author} {\bibfnamefont {I.}~\bibnamefont {Talmi}},\ }\href@noop {} {\emph {\bibinfo {title} {Nuclear shell theory}}},\ Vol.~\bibinfo {volume} {14}\ (\bibinfo  {publisher} {Academic Press},\ \bibinfo {year} {2013})\BibitemShut {NoStop}%
\bibitem [{\citenamefont {Qi}\ and\ \citenamefont {Zhang}(2010)}]{Qi2010T}%
  \BibitemOpen
  \bibfield  {author} {\bibinfo {author} {\bibfnamefont {X.-L.}\ \bibnamefont {Qi}}\ and\ \bibinfo {author} {\bibfnamefont {S.-C.}\ \bibnamefont {Zhang}},\ }\bibfield  {title} {\bibinfo {title} {The quantum spin hall effect and topological insulators},\ }\href {https://doi.org/10.1063/1.3293411} {\bibfield  {journal} {\bibinfo  {journal} {Physics Today}\ }\textbf {\bibinfo {volume} {63}},\ \bibinfo {pages} {33} (\bibinfo {year} {2010})}\BibitemShut {NoStop}%
\bibitem [{\citenamefont {Hasan}\ and\ \citenamefont {Kane}(2010)}]{hasan2010C}%
  \BibitemOpen
  \bibfield  {author} {\bibinfo {author} {\bibfnamefont {M.~Z.}\ \bibnamefont {Hasan}}\ and\ \bibinfo {author} {\bibfnamefont {C.~L.}\ \bibnamefont {Kane}},\ }\bibfield  {title} {\bibinfo {title} {Colloquium: Topological insulators},\ }\href {https://doi.org/10.1103/RevModPhys.82.3045} {\bibfield  {journal} {\bibinfo  {journal} {Rev. Mod. Phys.}\ }\textbf {\bibinfo {volume} {82}},\ \bibinfo {pages} {3045} (\bibinfo {year} {2010})}\BibitemShut {NoStop}%
\bibitem [{\citenamefont {Lin}\ \emph {et~al.}(2011)\citenamefont {Lin}, \citenamefont {Jimenez-Garcia},\ and\ \citenamefont {Spielman}}]{lin2011spin}%
  \BibitemOpen
  \bibfield  {author} {\bibinfo {author} {\bibfnamefont {Y.-J.}\ \bibnamefont {Lin}}, \bibinfo {author} {\bibfnamefont {K.}~\bibnamefont {Jimenez-Garcia}},\ and\ \bibinfo {author} {\bibfnamefont {I.~B.}\ \bibnamefont {Spielman}},\ }\bibfield  {title} {\bibinfo {title} {Spin-orbit-coupled bose-einstein condensates},\ }\href {https://doi.org/10.1038/nature09887} {\bibfield  {journal} {\bibinfo  {journal} {Nature (London)}\ }\textbf {\bibinfo {volume} {471}},\ \bibinfo {pages} {83} (\bibinfo {year} {2011})}\BibitemShut {NoStop}%
\bibitem [{\citenamefont {Wang}\ \emph {et~al.}(2012)\citenamefont {Wang}, \citenamefont {Yu}, \citenamefont {Fu}, \citenamefont {Miao}, \citenamefont {Huang}, \citenamefont {Chai}, \citenamefont {Zhai},\ and\ \citenamefont {Zhang}}]{wang2012spin}%
  \BibitemOpen
  \bibfield  {author} {\bibinfo {author} {\bibfnamefont {P.}~\bibnamefont {Wang}}, \bibinfo {author} {\bibfnamefont {Z.-Q.}\ \bibnamefont {Yu}}, \bibinfo {author} {\bibfnamefont {Z.}~\bibnamefont {Fu}}, \bibinfo {author} {\bibfnamefont {J.}~\bibnamefont {Miao}}, \bibinfo {author} {\bibfnamefont {L.}~\bibnamefont {Huang}}, \bibinfo {author} {\bibfnamefont {S.}~\bibnamefont {Chai}}, \bibinfo {author} {\bibfnamefont {H.}~\bibnamefont {Zhai}},\ and\ \bibinfo {author} {\bibfnamefont {J.}~\bibnamefont {Zhang}},\ }\bibfield  {title} {\bibinfo {title} {Spin-orbit coupled degenerate fermi gases},\ }\href {https://doi.org/10.1103/PhysRevLett.109.095301} {\bibfield  {journal} {\bibinfo  {journal} {Phys. Rev. Lett.}\ }\textbf {\bibinfo {volume} {109}},\ \bibinfo {pages} {095301} (\bibinfo {year} {2012})}\BibitemShut {NoStop}%
\bibitem [{\citenamefont {Cheuk}\ \emph {et~al.}(2012)\citenamefont {Cheuk}, \citenamefont {Sommer}, \citenamefont {Hadzibabic}, \citenamefont {Yefsah}, \citenamefont {Bakr},\ and\ \citenamefont {Zwierlein}}]{cheuk2012spin}%
  \BibitemOpen
  \bibfield  {author} {\bibinfo {author} {\bibfnamefont {L.~W.}\ \bibnamefont {Cheuk}}, \bibinfo {author} {\bibfnamefont {A.~T.}\ \bibnamefont {Sommer}}, \bibinfo {author} {\bibfnamefont {Z.}~\bibnamefont {Hadzibabic}}, \bibinfo {author} {\bibfnamefont {T.}~\bibnamefont {Yefsah}}, \bibinfo {author} {\bibfnamefont {W.~S.}\ \bibnamefont {Bakr}},\ and\ \bibinfo {author} {\bibfnamefont {M.~W.}\ \bibnamefont {Zwierlein}},\ }\bibfield  {title} {\bibinfo {title} {Spin-injection spectroscopy of a spin-orbit coupled fermi gas},\ }\href {https://doi.org/10.1103/PhysRevLett.109.095302} {\bibfield  {journal} {\bibinfo  {journal} {Phys. Rev. Lett.}\ }\textbf {\bibinfo {volume} {109}},\ \bibinfo {pages} {095302} (\bibinfo {year} {2012})}\BibitemShut {NoStop}%
\bibitem [{\citenamefont {Liu}\ \emph {et~al.}(2009)\citenamefont {Liu}, \citenamefont {Borunda}, \citenamefont {Liu},\ and\ \citenamefont {Sinova}}]{liu2009effect}%
  \BibitemOpen
  \bibfield  {author} {\bibinfo {author} {\bibfnamefont {X.-J.}\ \bibnamefont {Liu}}, \bibinfo {author} {\bibfnamefont {M.~F.}\ \bibnamefont {Borunda}}, \bibinfo {author} {\bibfnamefont {X.}~\bibnamefont {Liu}},\ and\ \bibinfo {author} {\bibfnamefont {J.}~\bibnamefont {Sinova}},\ }\bibfield  {title} {\bibinfo {title} {Effect of induced spin-orbit coupling for atoms via laser fields},\ }\href {https://doi.org/10.1103/PhysRevLett.102.046402} {\bibfield  {journal} {\bibinfo  {journal} {Phys. Rev. Lett.}\ }\textbf {\bibinfo {volume} {102}},\ \bibinfo {pages} {046402} (\bibinfo {year} {2009})}\BibitemShut {NoStop}%
\bibitem [{\citenamefont {Williams}\ \emph {et~al.}(2012)\citenamefont {Williams}, \citenamefont {LeBlanc}, \citenamefont {Jiménez-García}, \citenamefont {Beeler}, \citenamefont {Perry}, \citenamefont {Phillips},\ and\ \citenamefont {Spielman}}]{williams2012synthetic}%
  \BibitemOpen
  \bibfield  {author} {\bibinfo {author} {\bibfnamefont {R.~A.}\ \bibnamefont {Williams}}, \bibinfo {author} {\bibfnamefont {L.~J.}\ \bibnamefont {LeBlanc}}, \bibinfo {author} {\bibfnamefont {K.}~\bibnamefont {Jiménez-García}}, \bibinfo {author} {\bibfnamefont {M.~C.}\ \bibnamefont {Beeler}}, \bibinfo {author} {\bibfnamefont {A.~R.}\ \bibnamefont {Perry}}, \bibinfo {author} {\bibfnamefont {W.~D.}\ \bibnamefont {Phillips}},\ and\ \bibinfo {author} {\bibfnamefont {I.~B.}\ \bibnamefont {Spielman}},\ }\bibfield  {title} {\bibinfo {title} {Synthetic partial waves in ultracold atomic collisions},\ }\href {https://doi.org/10.1126/science.1212652} {\bibfield  {journal} {\bibinfo  {journal} {Science}\ }\textbf {\bibinfo {volume} {335}},\ \bibinfo {pages} {314} (\bibinfo {year} {2012})}\BibitemShut {NoStop}%
\bibitem [{\citenamefont {Zhang}\ \emph {et~al.}(2012)\citenamefont {Zhang}, \citenamefont {Ji}, \citenamefont {Chen}, \citenamefont {Zhang}, \citenamefont {Du}, \citenamefont {Yan}, \citenamefont {Pan}, \citenamefont {Zhao}, \citenamefont {Deng}, \citenamefont {Zhai}, \citenamefont {Chen},\ and\ \citenamefont {Pan}}]{zhang2012collective}%
  \BibitemOpen
  \bibfield  {author} {\bibinfo {author} {\bibfnamefont {J.-Y.}\ \bibnamefont {Zhang}}, \bibinfo {author} {\bibfnamefont {S.-C.}\ \bibnamefont {Ji}}, \bibinfo {author} {\bibfnamefont {Z.}~\bibnamefont {Chen}}, \bibinfo {author} {\bibfnamefont {L.}~\bibnamefont {Zhang}}, \bibinfo {author} {\bibfnamefont {Z.-D.}\ \bibnamefont {Du}}, \bibinfo {author} {\bibfnamefont {B.}~\bibnamefont {Yan}}, \bibinfo {author} {\bibfnamefont {G.-S.}\ \bibnamefont {Pan}}, \bibinfo {author} {\bibfnamefont {B.}~\bibnamefont {Zhao}}, \bibinfo {author} {\bibfnamefont {Y.-J.}\ \bibnamefont {Deng}}, \bibinfo {author} {\bibfnamefont {H.}~\bibnamefont {Zhai}}, \bibinfo {author} {\bibfnamefont {S.}~\bibnamefont {Chen}},\ and\ \bibinfo {author} {\bibfnamefont {J.-W.}\ \bibnamefont {Pan}},\ }\bibfield  {title} {\bibinfo {title} {Collective dipole oscillations of a spin-orbit coupled bose-einstein condensate},\ }\href {https://doi.org/10.1103/PhysRevLett.109.115301} {\bibfield  {journal} {\bibinfo  {journal} {Phys. Rev. Lett.}\ }\textbf
  {\bibinfo {volume} {109}},\ \bibinfo {pages} {115301} (\bibinfo {year} {2012})}\BibitemShut {NoStop}%
\bibitem [{\citenamefont {Olson}\ \emph {et~al.}(2014)\citenamefont {Olson}, \citenamefont {Wang}, \citenamefont {Niffenegger}, \citenamefont {Li}, \citenamefont {Greene},\ and\ \citenamefont {Chen}}]{olson2014tunable}%
  \BibitemOpen
  \bibfield  {author} {\bibinfo {author} {\bibfnamefont {A.~J.}\ \bibnamefont {Olson}}, \bibinfo {author} {\bibfnamefont {S.-J.}\ \bibnamefont {Wang}}, \bibinfo {author} {\bibfnamefont {R.~J.}\ \bibnamefont {Niffenegger}}, \bibinfo {author} {\bibfnamefont {C.-H.}\ \bibnamefont {Li}}, \bibinfo {author} {\bibfnamefont {C.~H.}\ \bibnamefont {Greene}},\ and\ \bibinfo {author} {\bibfnamefont {Y.~P.}\ \bibnamefont {Chen}},\ }\bibfield  {title} {\bibinfo {title} {Tunable landau-zener transitions in a spin-orbit-coupled bose-einstein condensate},\ }\href {https://doi.org/10.1103/PhysRevA.90.013616} {\bibfield  {journal} {\bibinfo  {journal} {Phys. Rev. A}\ }\textbf {\bibinfo {volume} {90}},\ \bibinfo {pages} {013616} (\bibinfo {year} {2014})}\BibitemShut {NoStop}%
\bibitem [{\citenamefont {Fu}\ \emph {et~al.}(2014)\citenamefont {Fu}, \citenamefont {Huang}, \citenamefont {Meng}, \citenamefont {Wang}, \citenamefont {Zhang}, \citenamefont {Zhang}, \citenamefont {Zhai}, \citenamefont {Zhang},\ and\ \citenamefont {Zhang}}]{fu2014production}%
  \BibitemOpen
  \bibfield  {author} {\bibinfo {author} {\bibfnamefont {Z.}~\bibnamefont {Fu}}, \bibinfo {author} {\bibfnamefont {L.}~\bibnamefont {Huang}}, \bibinfo {author} {\bibfnamefont {Z.}~\bibnamefont {Meng}}, \bibinfo {author} {\bibfnamefont {P.}~\bibnamefont {Wang}}, \bibinfo {author} {\bibfnamefont {L.}~\bibnamefont {Zhang}}, \bibinfo {author} {\bibfnamefont {S.}~\bibnamefont {Zhang}}, \bibinfo {author} {\bibfnamefont {H.}~\bibnamefont {Zhai}}, \bibinfo {author} {\bibfnamefont {P.}~\bibnamefont {Zhang}},\ and\ \bibinfo {author} {\bibfnamefont {J.}~\bibnamefont {Zhang}},\ }\bibfield  {title} {\bibinfo {title} {Production of feshbach molecules induced by spin--orbit coupling in fermi gases},\ }\href {https://doi.org/10.1038/nphys2824} {\bibfield  {journal} {\bibinfo  {journal} {Nature Physics}\ }\textbf {\bibinfo {volume} {10}},\ \bibinfo {pages} {110} (\bibinfo {year} {2014})}\BibitemShut {NoStop}%
\bibitem [{\citenamefont {Ji}\ \emph {et~al.}(2014)\citenamefont {Ji}, \citenamefont {Zhang}, \citenamefont {Zhang}, \citenamefont {Du}, \citenamefont {Zheng}, \citenamefont {Deng}, \citenamefont {Zhai}, \citenamefont {Chen},\ and\ \citenamefont {Pan}}]{ji2014experimental}%
  \BibitemOpen
  \bibfield  {author} {\bibinfo {author} {\bibfnamefont {S.-C.}\ \bibnamefont {Ji}}, \bibinfo {author} {\bibfnamefont {J.-Y.}\ \bibnamefont {Zhang}}, \bibinfo {author} {\bibfnamefont {L.}~\bibnamefont {Zhang}}, \bibinfo {author} {\bibfnamefont {Z.-D.}\ \bibnamefont {Du}}, \bibinfo {author} {\bibfnamefont {W.}~\bibnamefont {Zheng}}, \bibinfo {author} {\bibfnamefont {Y.-J.}\ \bibnamefont {Deng}}, \bibinfo {author} {\bibfnamefont {H.}~\bibnamefont {Zhai}}, \bibinfo {author} {\bibfnamefont {S.}~\bibnamefont {Chen}},\ and\ \bibinfo {author} {\bibfnamefont {J.-W.}\ \bibnamefont {Pan}},\ }\bibfield  {title} {\bibinfo {title} {Experimental determination of the finite-temperature phase diagram of a spin-orbit coupled bose gas},\ }\href {http://dx.doi.org/10.1038/nphys2905} {\bibfield  {journal} {\bibinfo  {journal} {Nat Phys}\ }\textbf {\bibinfo {volume} {10}},\ \bibinfo {pages} {314} (\bibinfo {year} {2014})}\BibitemShut {NoStop}%
\bibitem [{\citenamefont {Ji}\ \emph {et~al.}(2015)\citenamefont {Ji}, \citenamefont {Zhang}, \citenamefont {Xu}, \citenamefont {Wu}, \citenamefont {Deng}, \citenamefont {Chen},\ and\ \citenamefont {Pan}}]{ji2015softening}%
  \BibitemOpen
  \bibfield  {author} {\bibinfo {author} {\bibfnamefont {S.-C.}\ \bibnamefont {Ji}}, \bibinfo {author} {\bibfnamefont {L.}~\bibnamefont {Zhang}}, \bibinfo {author} {\bibfnamefont {X.-T.}\ \bibnamefont {Xu}}, \bibinfo {author} {\bibfnamefont {Z.}~\bibnamefont {Wu}}, \bibinfo {author} {\bibfnamefont {Y.}~\bibnamefont {Deng}}, \bibinfo {author} {\bibfnamefont {S.}~\bibnamefont {Chen}},\ and\ \bibinfo {author} {\bibfnamefont {J.-W.}\ \bibnamefont {Pan}},\ }\bibfield  {title} {\bibinfo {title} {Softening of roton and phonon modes in a bose-einstein condensate with spin-orbit coupling},\ }\href {https://doi.org/10.1103/PhysRevLett.114.105301} {\bibfield  {journal} {\bibinfo  {journal} {Phys. Rev. Lett.}\ }\textbf {\bibinfo {volume} {114}},\ \bibinfo {pages} {105301} (\bibinfo {year} {2015})}\BibitemShut {NoStop}%
\bibitem [{\citenamefont {Hamner}\ \emph {et~al.}(2015)\citenamefont {Hamner}, \citenamefont {Zhang}, \citenamefont {Khamehchi}, \citenamefont {Davis},\ and\ \citenamefont {Engels}}]{hamner2015spin}%
  \BibitemOpen
  \bibfield  {author} {\bibinfo {author} {\bibfnamefont {C.}~\bibnamefont {Hamner}}, \bibinfo {author} {\bibfnamefont {Y.}~\bibnamefont {Zhang}}, \bibinfo {author} {\bibfnamefont {M.~A.}\ \bibnamefont {Khamehchi}}, \bibinfo {author} {\bibfnamefont {M.~J.}\ \bibnamefont {Davis}},\ and\ \bibinfo {author} {\bibfnamefont {P.}~\bibnamefont {Engels}},\ }\bibfield  {title} {\bibinfo {title} {Spin-orbit-coupled bose-einstein condensates in a one-dimensional optical lattice},\ }\href {https://doi.org/10.1103/PhysRevLett.114.070401} {\bibfield  {journal} {\bibinfo  {journal} {Phys. Rev. Lett.}\ }\textbf {\bibinfo {volume} {114}},\ \bibinfo {pages} {070401} (\bibinfo {year} {2015})}\BibitemShut {NoStop}%
\bibitem [{\citenamefont {Jim\'enez-Garc\'{\i}a}\ \emph {et~al.}(2015)\citenamefont {Jim\'enez-Garc\'{\i}a}, \citenamefont {LeBlanc}, \citenamefont {Williams}, \citenamefont {Beeler}, \citenamefont {Qu}, \citenamefont {Gong}, \citenamefont {Zhang},\ and\ \citenamefont {Spielman}}]{garcia2015tunable}%
  \BibitemOpen
  \bibfield  {author} {\bibinfo {author} {\bibfnamefont {K.}~\bibnamefont {Jim\'enez-Garc\'{\i}a}}, \bibinfo {author} {\bibfnamefont {L.~J.}\ \bibnamefont {LeBlanc}}, \bibinfo {author} {\bibfnamefont {R.~A.}\ \bibnamefont {Williams}}, \bibinfo {author} {\bibfnamefont {M.~C.}\ \bibnamefont {Beeler}}, \bibinfo {author} {\bibfnamefont {C.}~\bibnamefont {Qu}}, \bibinfo {author} {\bibfnamefont {M.}~\bibnamefont {Gong}}, \bibinfo {author} {\bibfnamefont {C.}~\bibnamefont {Zhang}},\ and\ \bibinfo {author} {\bibfnamefont {I.~B.}\ \bibnamefont {Spielman}},\ }\bibfield  {title} {\bibinfo {title} {Tunable spin-orbit coupling via strong driving in ultracold-atom systems},\ }\href {https://doi.org/10.1103/PhysRevLett.114.125301} {\bibfield  {journal} {\bibinfo  {journal} {Phys. Rev. Lett.}\ }\textbf {\bibinfo {volume} {114}},\ \bibinfo {pages} {125301} (\bibinfo {year} {2015})}\BibitemShut {NoStop}%
\bibitem [{\citenamefont {Burdick}\ \emph {et~al.}(2016)\citenamefont {Burdick}, \citenamefont {Tang},\ and\ \citenamefont {Lev}}]{burdick2016long}%
  \BibitemOpen
  \bibfield  {author} {\bibinfo {author} {\bibfnamefont {N.~Q.}\ \bibnamefont {Burdick}}, \bibinfo {author} {\bibfnamefont {Y.}~\bibnamefont {Tang}},\ and\ \bibinfo {author} {\bibfnamefont {B.~L.}\ \bibnamefont {Lev}},\ }\bibfield  {title} {\bibinfo {title} {Long-lived spin-orbit-coupled degenerate dipolar fermi gas},\ }\href {https://doi.org/10.1103/PhysRevX.6.031022} {\bibfield  {journal} {\bibinfo  {journal} {Phys. Rev. X}\ }\textbf {\bibinfo {volume} {6}},\ \bibinfo {pages} {031022} (\bibinfo {year} {2016})}\BibitemShut {NoStop}%
\bibitem [{\citenamefont {Song}\ \emph {et~al.}(2016)\citenamefont {Song}, \citenamefont {He}, \citenamefont {Zhang}, \citenamefont {Hajiyev}, \citenamefont {Huang}, \citenamefont {Liu},\ and\ \citenamefont {Jo}}]{song2016spin}%
  \BibitemOpen
  \bibfield  {author} {\bibinfo {author} {\bibfnamefont {B.}~\bibnamefont {Song}}, \bibinfo {author} {\bibfnamefont {C.}~\bibnamefont {He}}, \bibinfo {author} {\bibfnamefont {S.}~\bibnamefont {Zhang}}, \bibinfo {author} {\bibfnamefont {E.}~\bibnamefont {Hajiyev}}, \bibinfo {author} {\bibfnamefont {W.}~\bibnamefont {Huang}}, \bibinfo {author} {\bibfnamefont {X.-J.}\ \bibnamefont {Liu}},\ and\ \bibinfo {author} {\bibfnamefont {G.-B.}\ \bibnamefont {Jo}},\ }\bibfield  {title} {\bibinfo {title} {Spin-orbit-coupled two-electron fermi gases of ytterbium atoms},\ }\href {https://doi.org/10.1103/PhysRevA.94.061604} {\bibfield  {journal} {\bibinfo  {journal} {Phys. Rev. A}\ }\textbf {\bibinfo {volume} {94}},\ \bibinfo {pages} {061604} (\bibinfo {year} {2016})}\BibitemShut {NoStop}%
\bibitem [{\citenamefont {Li}\ \emph {et~al.}(2016)\citenamefont {Li}, \citenamefont {Huang}, \citenamefont {Shteynas}, \citenamefont {Burchesky}, \citenamefont {Top}, \citenamefont {Su}, \citenamefont {Lee}, \citenamefont {Jamison},\ and\ \citenamefont {Ketterle}}]{li2016spin}%
  \BibitemOpen
  \bibfield  {author} {\bibinfo {author} {\bibfnamefont {J.}~\bibnamefont {Li}}, \bibinfo {author} {\bibfnamefont {W.}~\bibnamefont {Huang}}, \bibinfo {author} {\bibfnamefont {B.}~\bibnamefont {Shteynas}}, \bibinfo {author} {\bibfnamefont {S.}~\bibnamefont {Burchesky}}, \bibinfo {author} {\bibfnamefont {F.~i. m. c. b. u. i. e. i.~f.}\ \bibnamefont {Top}}, \bibinfo {author} {\bibfnamefont {E.}~\bibnamefont {Su}}, \bibinfo {author} {\bibfnamefont {J.}~\bibnamefont {Lee}}, \bibinfo {author} {\bibfnamefont {A.~O.}\ \bibnamefont {Jamison}},\ and\ \bibinfo {author} {\bibfnamefont {W.}~\bibnamefont {Ketterle}},\ }\bibfield  {title} {\bibinfo {title} {Spin-orbit coupling and spin textures in optical superlattices},\ }\href {https://doi.org/10.1103/PhysRevLett.117.185301} {\bibfield  {journal} {\bibinfo  {journal} {Phys. Rev. Lett.}\ }\textbf {\bibinfo {volume} {117}},\ \bibinfo {pages} {185301} (\bibinfo {year} {2016})}\BibitemShut {NoStop}%
\bibitem [{\citenamefont {Livi}\ \emph {et~al.}(2016)\citenamefont {Livi}, \citenamefont {Cappellini}, \citenamefont {Diem}, \citenamefont {Franchi}, \citenamefont {Clivati}, \citenamefont {Frittelli}, \citenamefont {Levi}, \citenamefont {Calonico}, \citenamefont {Catani}, \citenamefont {Inguscio},\ and\ \citenamefont {Fallani}}]{livi2016synthetic}%
  \BibitemOpen
  \bibfield  {author} {\bibinfo {author} {\bibfnamefont {L.~F.}\ \bibnamefont {Livi}}, \bibinfo {author} {\bibfnamefont {G.}~\bibnamefont {Cappellini}}, \bibinfo {author} {\bibfnamefont {M.}~\bibnamefont {Diem}}, \bibinfo {author} {\bibfnamefont {L.}~\bibnamefont {Franchi}}, \bibinfo {author} {\bibfnamefont {C.}~\bibnamefont {Clivati}}, \bibinfo {author} {\bibfnamefont {M.}~\bibnamefont {Frittelli}}, \bibinfo {author} {\bibfnamefont {F.}~\bibnamefont {Levi}}, \bibinfo {author} {\bibfnamefont {D.}~\bibnamefont {Calonico}}, \bibinfo {author} {\bibfnamefont {J.}~\bibnamefont {Catani}}, \bibinfo {author} {\bibfnamefont {M.}~\bibnamefont {Inguscio}},\ and\ \bibinfo {author} {\bibfnamefont {L.}~\bibnamefont {Fallani}},\ }\bibfield  {title} {\bibinfo {title} {Synthetic dimensions and spin-orbit coupling with an optical clock transition},\ }\href {https://doi.org/10.1103/PhysRevLett.117.220401} {\bibfield  {journal} {\bibinfo  {journal} {Phys. Rev. Lett.}\ }\textbf {\bibinfo {volume} {117}},\ \bibinfo {pages}
  {220401} (\bibinfo {year} {2016})}\BibitemShut {NoStop}%
\bibitem [{\citenamefont {Osterloh}\ \emph {et~al.}(2005)\citenamefont {Osterloh}, \citenamefont {Baig}, \citenamefont {Santos}, \citenamefont {Zoller},\ and\ \citenamefont {Lewenstein}}]{osterloh2005cold}%
  \BibitemOpen
  \bibfield  {author} {\bibinfo {author} {\bibfnamefont {K.}~\bibnamefont {Osterloh}}, \bibinfo {author} {\bibfnamefont {M.}~\bibnamefont {Baig}}, \bibinfo {author} {\bibfnamefont {L.}~\bibnamefont {Santos}}, \bibinfo {author} {\bibfnamefont {P.}~\bibnamefont {Zoller}},\ and\ \bibinfo {author} {\bibfnamefont {M.}~\bibnamefont {Lewenstein}},\ }\bibfield  {title} {\bibinfo {title} {Cold atoms in non-abelian gauge potentials: From the hofstadter "moth" to lattice gauge theory},\ }\href {https://doi.org/10.1103/PhysRevLett.95.010403} {\bibfield  {journal} {\bibinfo  {journal} {Phys. Rev. Lett.}\ }\textbf {\bibinfo {volume} {95}},\ \bibinfo {pages} {010403} (\bibinfo {year} {2005})}\BibitemShut {NoStop}%
\bibitem [{\citenamefont {Ruseckas}\ \emph {et~al.}(2005)\citenamefont {Ruseckas}, \citenamefont {Juzeli\ifmmode~\bar{u}\else \={u}\fi{}nas}, \citenamefont {\"Ohberg},\ and\ \citenamefont {Fleischhauer}}]{ruseckas2005non}%
  \BibitemOpen
  \bibfield  {author} {\bibinfo {author} {\bibfnamefont {J.}~\bibnamefont {Ruseckas}}, \bibinfo {author} {\bibfnamefont {G.}~\bibnamefont {Juzeli\ifmmode~\bar{u}\else \={u}\fi{}nas}}, \bibinfo {author} {\bibfnamefont {P.}~\bibnamefont {\"Ohberg}},\ and\ \bibinfo {author} {\bibfnamefont {M.}~\bibnamefont {Fleischhauer}},\ }\bibfield  {title} {\bibinfo {title} {Non-abelian gauge potentials for ultracold atoms with degenerate dark states},\ }\href {https://doi.org/10.1103/PhysRevLett.95.010404} {\bibfield  {journal} {\bibinfo  {journal} {Phys. Rev. Lett.}\ }\textbf {\bibinfo {volume} {95}},\ \bibinfo {pages} {010404} (\bibinfo {year} {2005})}\BibitemShut {NoStop}%
\bibitem [{\citenamefont {Juzeli\ifmmode~\bar{u}\else \={u}\fi{}nas}\ \emph {et~al.}(2010)\citenamefont {Juzeli\ifmmode~\bar{u}\else \={u}\fi{}nas}, \citenamefont {Ruseckas},\ and\ \citenamefont {Dalibard}}]{juzeliunas2010generalized}%
  \BibitemOpen
  \bibfield  {author} {\bibinfo {author} {\bibfnamefont {G.}~\bibnamefont {Juzeli\ifmmode~\bar{u}\else \={u}\fi{}nas}}, \bibinfo {author} {\bibfnamefont {J.}~\bibnamefont {Ruseckas}},\ and\ \bibinfo {author} {\bibfnamefont {J.}~\bibnamefont {Dalibard}},\ }\bibfield  {title} {\bibinfo {title} {Generalized rashba-dresselhaus spin-orbit coupling for cold atoms},\ }\href {https://doi.org/10.1103/PhysRevA.81.053403} {\bibfield  {journal} {\bibinfo  {journal} {Phys. Rev. A}\ }\textbf {\bibinfo {volume} {81}},\ \bibinfo {pages} {053403} (\bibinfo {year} {2010})}\BibitemShut {NoStop}%
\bibitem [{\citenamefont {Campbell}\ \emph {et~al.}(2011)\citenamefont {Campbell}, \citenamefont {Juzeli\ifmmode~\bar{u}\else \={u}\fi{}nas},\ and\ \citenamefont {Spielman}}]{campbell2011realistic}%
  \BibitemOpen
  \bibfield  {author} {\bibinfo {author} {\bibfnamefont {D.~L.}\ \bibnamefont {Campbell}}, \bibinfo {author} {\bibfnamefont {G.}~\bibnamefont {Juzeli\ifmmode~\bar{u}\else \={u}\fi{}nas}},\ and\ \bibinfo {author} {\bibfnamefont {I.~B.}\ \bibnamefont {Spielman}},\ }\bibfield  {title} {\bibinfo {title} {Realistic rashba and dresselhaus spin-orbit coupling for neutral atoms},\ }\href {https://doi.org/10.1103/PhysRevA.84.025602} {\bibfield  {journal} {\bibinfo  {journal} {Phys. Rev. A}\ }\textbf {\bibinfo {volume} {84}},\ \bibinfo {pages} {025602} (\bibinfo {year} {2011})}\BibitemShut {NoStop}%
\bibitem [{\citenamefont {Sau}\ \emph {et~al.}(2011)\citenamefont {Sau}, \citenamefont {Sensarma}, \citenamefont {Powell}, \citenamefont {Spielman},\ and\ \citenamefont {Das~Sarma}}]{sau2011chiral}%
  \BibitemOpen
  \bibfield  {author} {\bibinfo {author} {\bibfnamefont {J.~D.}\ \bibnamefont {Sau}}, \bibinfo {author} {\bibfnamefont {R.}~\bibnamefont {Sensarma}}, \bibinfo {author} {\bibfnamefont {S.}~\bibnamefont {Powell}}, \bibinfo {author} {\bibfnamefont {I.~B.}\ \bibnamefont {Spielman}},\ and\ \bibinfo {author} {\bibfnamefont {S.}~\bibnamefont {Das~Sarma}},\ }\bibfield  {title} {\bibinfo {title} {Chiral rashba spin textures in ultracold fermi gases},\ }\href {https://doi.org/10.1103/PhysRevB.83.140510} {\bibfield  {journal} {\bibinfo  {journal} {Phys. Rev. B}\ }\textbf {\bibinfo {volume} {83}},\ \bibinfo {pages} {140510} (\bibinfo {year} {2011})}\BibitemShut {NoStop}%
\bibitem [{\citenamefont {Anderson}\ \emph {et~al.}(2013)\citenamefont {Anderson}, \citenamefont {Spielman},\ and\ \citenamefont {Juzeli\ifmmode~\bar{u}\else \={u}\fi{}nas}}]{anderson2013magnetically}%
  \BibitemOpen
  \bibfield  {author} {\bibinfo {author} {\bibfnamefont {B.~M.}\ \bibnamefont {Anderson}}, \bibinfo {author} {\bibfnamefont {I.~B.}\ \bibnamefont {Spielman}},\ and\ \bibinfo {author} {\bibfnamefont {G.}~\bibnamefont {Juzeli\ifmmode~\bar{u}\else \={u}\fi{}nas}},\ }\bibfield  {title} {\bibinfo {title} {Magnetically generated spin-orbit coupling for ultracold atoms},\ }\href {https://doi.org/10.1103/PhysRevLett.111.125301} {\bibfield  {journal} {\bibinfo  {journal} {Phys. Rev. Lett.}\ }\textbf {\bibinfo {volume} {111}},\ \bibinfo {pages} {125301} (\bibinfo {year} {2013})}\BibitemShut {NoStop}%
\bibitem [{\citenamefont {Xu}\ \emph {et~al.}(2013)\citenamefont {Xu}, \citenamefont {You},\ and\ \citenamefont {Ueda}}]{xu2013atomic}%
  \BibitemOpen
  \bibfield  {author} {\bibinfo {author} {\bibfnamefont {Z.-F.}\ \bibnamefont {Xu}}, \bibinfo {author} {\bibfnamefont {L.}~\bibnamefont {You}},\ and\ \bibinfo {author} {\bibfnamefont {M.}~\bibnamefont {Ueda}},\ }\bibfield  {title} {\bibinfo {title} {Atomic spin-orbit coupling synthesized with magnetic-field-gradient pulses},\ }\href {https://doi.org/10.1103/PhysRevA.87.063634} {\bibfield  {journal} {\bibinfo  {journal} {Phys. Rev. A}\ }\textbf {\bibinfo {volume} {87}},\ \bibinfo {pages} {063634} (\bibinfo {year} {2013})}\BibitemShut {NoStop}%
\bibitem [{\citenamefont {Liu}\ \emph {et~al.}(2014)\citenamefont {Liu}, \citenamefont {Law},\ and\ \citenamefont {Ng}}]{liu2014realization}%
  \BibitemOpen
  \bibfield  {author} {\bibinfo {author} {\bibfnamefont {X.-J.}\ \bibnamefont {Liu}}, \bibinfo {author} {\bibfnamefont {K.~T.}\ \bibnamefont {Law}},\ and\ \bibinfo {author} {\bibfnamefont {T.~K.}\ \bibnamefont {Ng}},\ }\bibfield  {title} {\bibinfo {title} {Realization of 2d spin-orbit interaction and exotic topological orders in cold atoms},\ }\href {https://doi.org/10.1103/PhysRevLett.112.086401} {\bibfield  {journal} {\bibinfo  {journal} {Phys. Rev. Lett.}\ }\textbf {\bibinfo {volume} {112}},\ \bibinfo {pages} {086401} (\bibinfo {year} {2014})}\BibitemShut {NoStop}%
\bibitem [{\citenamefont {Anderson}\ \emph {et~al.}(2012)\citenamefont {Anderson}, \citenamefont {Juzeli\ifmmode~\bar{u}\else \={u}\fi{}nas}, \citenamefont {Galitski},\ and\ \citenamefont {Spielman}}]{anderson2012synthetic}%
  \BibitemOpen
  \bibfield  {author} {\bibinfo {author} {\bibfnamefont {B.~M.}\ \bibnamefont {Anderson}}, \bibinfo {author} {\bibfnamefont {G.}~\bibnamefont {Juzeli\ifmmode~\bar{u}\else \={u}\fi{}nas}}, \bibinfo {author} {\bibfnamefont {V.~M.}\ \bibnamefont {Galitski}},\ and\ \bibinfo {author} {\bibfnamefont {I.~B.}\ \bibnamefont {Spielman}},\ }\bibfield  {title} {\bibinfo {title} {Synthetic 3d spin-orbit coupling},\ }\href {https://doi.org/10.1103/PhysRevLett.108.235301} {\bibfield  {journal} {\bibinfo  {journal} {Phys. Rev. Lett.}\ }\textbf {\bibinfo {volume} {108}},\ \bibinfo {pages} {235301} (\bibinfo {year} {2012})}\BibitemShut {NoStop}%
\bibitem [{\citenamefont {Lu}\ \emph {et~al.}(2020)\citenamefont {Lu}, \citenamefont {Wang},\ and\ \citenamefont {Liu}}]{lu2020ideal}%
  \BibitemOpen
  \bibfield  {author} {\bibinfo {author} {\bibfnamefont {Y.-H.}\ \bibnamefont {Lu}}, \bibinfo {author} {\bibfnamefont {B.-Z.}\ \bibnamefont {Wang}},\ and\ \bibinfo {author} {\bibfnamefont {X.-J.}\ \bibnamefont {Liu}},\ }\bibfield  {title} {\bibinfo {title} {Ideal weyl semimetal with 3d spin-orbit coupled ultracold quantum gas},\ }\href {https://doi.org/https://doi.org/10.1016/j.scib.2020.09.036} {\bibfield  {journal} {\bibinfo  {journal} {Science Bulletin}\ }\textbf {\bibinfo {volume} {65}},\ \bibinfo {pages} {2080} (\bibinfo {year} {2020})}\BibitemShut {NoStop}%
\bibitem [{\citenamefont {Wang}\ \emph {et~al.}(2018)\citenamefont {Wang}, \citenamefont {Lu}, \citenamefont {Sun}, \citenamefont {Chen}, \citenamefont {Deng},\ and\ \citenamefont {Liu}}]{wang2018dirac}%
  \BibitemOpen
  \bibfield  {author} {\bibinfo {author} {\bibfnamefont {B.-Z.}\ \bibnamefont {Wang}}, \bibinfo {author} {\bibfnamefont {Y.-H.}\ \bibnamefont {Lu}}, \bibinfo {author} {\bibfnamefont {W.}~\bibnamefont {Sun}}, \bibinfo {author} {\bibfnamefont {S.}~\bibnamefont {Chen}}, \bibinfo {author} {\bibfnamefont {Y.}~\bibnamefont {Deng}},\ and\ \bibinfo {author} {\bibfnamefont {X.-J.}\ \bibnamefont {Liu}},\ }\bibfield  {title} {\bibinfo {title} {Dirac-, rashba-, and weyl-type spin-orbit couplings: Toward experimental realization in ultracold atoms},\ }\href {https://doi.org/10.1103/PhysRevA.97.011605} {\bibfield  {journal} {\bibinfo  {journal} {Phys. Rev. A}\ }\textbf {\bibinfo {volume} {97}},\ \bibinfo {pages} {011605} (\bibinfo {year} {2018})}\BibitemShut {NoStop}%
\bibitem [{\citenamefont {Huang}\ \emph {et~al.}(2016)\citenamefont {Huang}, \citenamefont {Meng}, \citenamefont {Wang}, \citenamefont {Peng}, \citenamefont {Zhang}, \citenamefont {Chen}, \citenamefont {Li}, \citenamefont {Zhou},\ and\ \citenamefont {Zhang}}]{huang2016experimental}%
  \BibitemOpen
  \bibfield  {author} {\bibinfo {author} {\bibfnamefont {L.}~\bibnamefont {Huang}}, \bibinfo {author} {\bibfnamefont {Z.}~\bibnamefont {Meng}}, \bibinfo {author} {\bibfnamefont {P.}~\bibnamefont {Wang}}, \bibinfo {author} {\bibfnamefont {P.}~\bibnamefont {Peng}}, \bibinfo {author} {\bibfnamefont {S.-L.}\ \bibnamefont {Zhang}}, \bibinfo {author} {\bibfnamefont {L.}~\bibnamefont {Chen}}, \bibinfo {author} {\bibfnamefont {D.}~\bibnamefont {Li}}, \bibinfo {author} {\bibfnamefont {Q.}~\bibnamefont {Zhou}},\ and\ \bibinfo {author} {\bibfnamefont {J.}~\bibnamefont {Zhang}},\ }\bibfield  {title} {\bibinfo {title} {Experimental realization of two-dimensional synthetic spin-orbit coupling in ultracold fermi gases},\ }\href {http://dx.doi.org/10.1038/nphys3672} {\bibfield  {journal} {\bibinfo  {journal} {Nat Phys}\ }\textbf {\bibinfo {volume} {12}},\ \bibinfo {pages} {540} (\bibinfo {year} {2016})}\BibitemShut {NoStop}%
\bibitem [{\citenamefont {Wu}\ \emph {et~al.}(2016)\citenamefont {Wu}, \citenamefont {Zhang}, \citenamefont {Sun}, \citenamefont {Xu}, \citenamefont {Wang}, \citenamefont {Ji}, \citenamefont {Deng}, \citenamefont {Chen}, \citenamefont {Liu},\ and\ \citenamefont {Pan}}]{wu2016realization}%
  \BibitemOpen
  \bibfield  {author} {\bibinfo {author} {\bibfnamefont {Z.}~\bibnamefont {Wu}}, \bibinfo {author} {\bibfnamefont {L.}~\bibnamefont {Zhang}}, \bibinfo {author} {\bibfnamefont {W.}~\bibnamefont {Sun}}, \bibinfo {author} {\bibfnamefont {X.-T.}\ \bibnamefont {Xu}}, \bibinfo {author} {\bibfnamefont {B.-Z.}\ \bibnamefont {Wang}}, \bibinfo {author} {\bibfnamefont {S.-C.}\ \bibnamefont {Ji}}, \bibinfo {author} {\bibfnamefont {Y.}~\bibnamefont {Deng}}, \bibinfo {author} {\bibfnamefont {S.}~\bibnamefont {Chen}}, \bibinfo {author} {\bibfnamefont {X.-J.}\ \bibnamefont {Liu}},\ and\ \bibinfo {author} {\bibfnamefont {J.-W.}\ \bibnamefont {Pan}},\ }\bibfield  {title} {\bibinfo {title} {Realization of two-dimensional spin-orbit coupling for bose-einstein condensates},\ }\href {https://doi.org/10.1126/science.aaf6689} {\bibfield  {journal} {\bibinfo  {journal} {Science}\ }\textbf {\bibinfo {volume} {354}},\ \bibinfo {pages} {83} (\bibinfo {year} {2016})}\BibitemShut {NoStop}%
\bibitem [{\citenamefont {Wang}\ \emph {et~al.}(2021{\natexlab{a}})\citenamefont {Wang}, \citenamefont {Cheng}, \citenamefont {Wang}, \citenamefont {Zhang}, \citenamefont {Lu}, \citenamefont {Yi}, \citenamefont {Niu}, \citenamefont {Deng}, \citenamefont {Liu}, \citenamefont {Chen} \emph {et~al.}}]{wang2021realization}%
  \BibitemOpen
  \bibfield  {author} {\bibinfo {author} {\bibfnamefont {Z.-Y.}\ \bibnamefont {Wang}}, \bibinfo {author} {\bibfnamefont {X.-C.}\ \bibnamefont {Cheng}}, \bibinfo {author} {\bibfnamefont {B.-Z.}\ \bibnamefont {Wang}}, \bibinfo {author} {\bibfnamefont {J.-Y.}\ \bibnamefont {Zhang}}, \bibinfo {author} {\bibfnamefont {Y.-H.}\ \bibnamefont {Lu}}, \bibinfo {author} {\bibfnamefont {C.-R.}\ \bibnamefont {Yi}}, \bibinfo {author} {\bibfnamefont {S.}~\bibnamefont {Niu}}, \bibinfo {author} {\bibfnamefont {Y.}~\bibnamefont {Deng}}, \bibinfo {author} {\bibfnamefont {X.-J.}\ \bibnamefont {Liu}}, \bibinfo {author} {\bibfnamefont {S.}~\bibnamefont {Chen}}, \emph {et~al.},\ }\bibfield  {title} {\bibinfo {title} {Realization of an ideal weyl semimetal band in a quantum gas with 3d spin-orbit coupling},\ }\href {https://doi.org/10.1126/science.abc0105} {\bibfield  {journal} {\bibinfo  {journal} {Science}\ }\textbf {\bibinfo {volume} {372}},\ \bibinfo {pages} {271} (\bibinfo {year} {2021}{\natexlab{a}})}\BibitemShut {NoStop}%
\bibitem [{\citenamefont {Galitski}\ and\ \citenamefont {Spielman}(2013)}]{galitski2013spin}%
  \BibitemOpen
  \bibfield  {author} {\bibinfo {author} {\bibfnamefont {V.}~\bibnamefont {Galitski}}\ and\ \bibinfo {author} {\bibfnamefont {I.~B.}\ \bibnamefont {Spielman}},\ }\bibfield  {title} {\bibinfo {title} {Spin--orbit coupling in quantum gases},\ }\href {https://doi.org/10.1038/nature11841} {\bibfield  {journal} {\bibinfo  {journal} {Nature}\ }\textbf {\bibinfo {volume} {494}},\ \bibinfo {pages} {49} (\bibinfo {year} {2013})}\BibitemShut {NoStop}%
\bibitem [{\citenamefont {Zhai}(2015)}]{zhai2015degenerate}%
  \BibitemOpen
  \bibfield  {author} {\bibinfo {author} {\bibfnamefont {H.}~\bibnamefont {Zhai}},\ }\bibfield  {title} {\bibinfo {title} {Degenerate quantum gases with spin–orbit coupling: a review},\ }\href {https://doi.org/10.1088/0034-4885/78/2/026001} {\bibfield  {journal} {\bibinfo  {journal} {Reports on Progress in Physics}\ }\textbf {\bibinfo {volume} {78}},\ \bibinfo {pages} {026001} (\bibinfo {year} {2015})}\BibitemShut {NoStop}%
\bibitem [{\citenamefont {Zhang}\ and\ \citenamefont {Liu}(2018)}]{Zhang2018S}%
  \BibitemOpen
  \bibfield  {author} {\bibinfo {author} {\bibfnamefont {L.}~\bibnamefont {Zhang}}\ and\ \bibinfo {author} {\bibfnamefont {X.-J.}\ \bibnamefont {Liu}},\ }\bibfield  {title} {\bibinfo {title} {Spin-orbit coupling and topological phases for ultracold atoms},\ }in\ \href {https://doi.org/10.1142/9789813272538_0001} {\emph {\bibinfo {booktitle} {Synthetic Spin-Orbit Coupling in Cold Atoms}}}\ (\bibinfo  {publisher} {World Scientific},\ \bibinfo {year} {2018})\ pp.\ \bibinfo {pages} {1--87}\BibitemShut {NoStop}%
\bibitem [{\citenamefont {Liu}\ \emph {et~al.}(2006)\citenamefont {Liu}, \citenamefont {Jing}, \citenamefont {Liu},\ and\ \citenamefont {Ge}}]{Liu2006G}%
  \BibitemOpen
  \bibfield  {author} {\bibinfo {author} {\bibfnamefont {X.~J.}\ \bibnamefont {Liu}}, \bibinfo {author} {\bibfnamefont {H.}~\bibnamefont {Jing}}, \bibinfo {author} {\bibfnamefont {X.}~\bibnamefont {Liu}},\ and\ \bibinfo {author} {\bibfnamefont {M.~L.}\ \bibnamefont {Ge}},\ }\bibfield  {title} {\bibinfo {title} {Generation of two-flavor vortex atom laser from a five-state medium},\ }\href {https://doi.org/10.1140/epjd/e2005-00260-0} {\bibfield  {journal} {\bibinfo  {journal} {The European Physical Journal D - Atomic, Molecular, Optical and Plasma Physics}\ }\textbf {\bibinfo {volume} {37}},\ \bibinfo {pages} {261} (\bibinfo {year} {2006})}\BibitemShut {NoStop}%
\bibitem [{\citenamefont {DeMarco}\ and\ \citenamefont {Pu}(2015)}]{demarco2015angular}%
  \BibitemOpen
  \bibfield  {author} {\bibinfo {author} {\bibfnamefont {M.}~\bibnamefont {DeMarco}}\ and\ \bibinfo {author} {\bibfnamefont {H.}~\bibnamefont {Pu}},\ }\bibfield  {title} {\bibinfo {title} {Angular spin-orbit coupling in cold atoms},\ }\href {https://doi.org/10.1103/PhysRevA.91.033630} {\bibfield  {journal} {\bibinfo  {journal} {Phys. Rev. A}\ }\textbf {\bibinfo {volume} {91}},\ \bibinfo {pages} {033630} (\bibinfo {year} {2015})}\BibitemShut {NoStop}%
\bibitem [{\citenamefont {Sun}\ \emph {et~al.}(2015)\citenamefont {Sun}, \citenamefont {Qu},\ and\ \citenamefont {Zhang}}]{sun2015spin}%
  \BibitemOpen
  \bibfield  {author} {\bibinfo {author} {\bibfnamefont {K.}~\bibnamefont {Sun}}, \bibinfo {author} {\bibfnamefont {C.}~\bibnamefont {Qu}},\ and\ \bibinfo {author} {\bibfnamefont {C.}~\bibnamefont {Zhang}},\ }\bibfield  {title} {\bibinfo {title} {Spin--orbital-angular-momentum coupling in bose-einstein condensates},\ }\href {https://doi.org/10.1103/PhysRevA.91.063627} {\bibfield  {journal} {\bibinfo  {journal} {Phys. Rev. A}\ }\textbf {\bibinfo {volume} {91}},\ \bibinfo {pages} {063627} (\bibinfo {year} {2015})}\BibitemShut {NoStop}%
\bibitem [{\citenamefont {Qu}\ \emph {et~al.}(2015)\citenamefont {Qu}, \citenamefont {Sun},\ and\ \citenamefont {Zhang}}]{qu2015quantum}%
  \BibitemOpen
  \bibfield  {author} {\bibinfo {author} {\bibfnamefont {C.}~\bibnamefont {Qu}}, \bibinfo {author} {\bibfnamefont {K.}~\bibnamefont {Sun}},\ and\ \bibinfo {author} {\bibfnamefont {C.}~\bibnamefont {Zhang}},\ }\bibfield  {title} {\bibinfo {title} {Quantum phases of bose-einstein condensates with synthetic spin--orbital-angular-momentum coupling},\ }\href {https://doi.org/10.1103/PhysRevA.91.053630} {\bibfield  {journal} {\bibinfo  {journal} {Phys. Rev. A}\ }\textbf {\bibinfo {volume} {91}},\ \bibinfo {pages} {053630} (\bibinfo {year} {2015})}\BibitemShut {NoStop}%
\bibitem [{\citenamefont {Hu}\ \emph {et~al.}(2015)\citenamefont {Hu}, \citenamefont {Miniatura},\ and\ \citenamefont {Gr\'emaud}}]{hu2015half}%
  \BibitemOpen
  \bibfield  {author} {\bibinfo {author} {\bibfnamefont {Y.-X.}\ \bibnamefont {Hu}}, \bibinfo {author} {\bibfnamefont {C.}~\bibnamefont {Miniatura}},\ and\ \bibinfo {author} {\bibfnamefont {B.}~\bibnamefont {Gr\'emaud}},\ }\bibfield  {title} {\bibinfo {title} {Half-skyrmion and vortex-antivortex pairs in spinor condensates},\ }\href {https://doi.org/10.1103/PhysRevA.92.033615} {\bibfield  {journal} {\bibinfo  {journal} {Phys. Rev. A}\ }\textbf {\bibinfo {volume} {92}},\ \bibinfo {pages} {033615} (\bibinfo {year} {2015})}\BibitemShut {NoStop}%
\bibitem [{\citenamefont {Chen}\ \emph {et~al.}(2016)\citenamefont {Chen}, \citenamefont {Pu},\ and\ \citenamefont {Zhang}}]{chen2016spin}%
  \BibitemOpen
  \bibfield  {author} {\bibinfo {author} {\bibfnamefont {L.}~\bibnamefont {Chen}}, \bibinfo {author} {\bibfnamefont {H.}~\bibnamefont {Pu}},\ and\ \bibinfo {author} {\bibfnamefont {Y.}~\bibnamefont {Zhang}},\ }\bibfield  {title} {\bibinfo {title} {Spin-orbit angular momentum coupling in a spin-1 bose-einstein condensate},\ }\href {https://doi.org/10.1103/PhysRevA.93.013629} {\bibfield  {journal} {\bibinfo  {journal} {Phys. Rev. A}\ }\textbf {\bibinfo {volume} {93}},\ \bibinfo {pages} {013629} (\bibinfo {year} {2016})}\BibitemShut {NoStop}%
\bibitem [{\citenamefont {Vasi\ifmmode~\acute{c}\else \'{c}\fi{}}\ and\ \citenamefont {Bala\ifmmode~\check{z}\else \v{z}\fi{}}(2016)}]{Vasic2016E}%
  \BibitemOpen
  \bibfield  {author} {\bibinfo {author} {\bibfnamefont {I.}~\bibnamefont {Vasi\ifmmode~\acute{c}\else \'{c}\fi{}}}\ and\ \bibinfo {author} {\bibfnamefont {A.}~\bibnamefont {Bala\ifmmode~\check{z}\else \v{z}\fi{}}},\ }\bibfield  {title} {\bibinfo {title} {Excitation spectra of a bose-einstein condensate with an angular spin-orbit coupling},\ }\href {https://doi.org/10.1103/PhysRevA.94.033627} {\bibfield  {journal} {\bibinfo  {journal} {Phys. Rev. A}\ }\textbf {\bibinfo {volume} {94}},\ \bibinfo {pages} {033627} (\bibinfo {year} {2016})}\BibitemShut {NoStop}%
\bibitem [{\citenamefont {Hou}\ \emph {et~al.}(2017)\citenamefont {Hou}, \citenamefont {Luo}, \citenamefont {Sun},\ and\ \citenamefont {Zhang}}]{Hou2017A}%
  \BibitemOpen
  \bibfield  {author} {\bibinfo {author} {\bibfnamefont {J.}~\bibnamefont {Hou}}, \bibinfo {author} {\bibfnamefont {X.-W.}\ \bibnamefont {Luo}}, \bibinfo {author} {\bibfnamefont {K.}~\bibnamefont {Sun}},\ and\ \bibinfo {author} {\bibfnamefont {C.}~\bibnamefont {Zhang}},\ }\bibfield  {title} {\bibinfo {title} {Adiabatically tuning quantized supercurrents in an annular bose-einstein condensate},\ }\href {https://doi.org/10.1103/PhysRevA.96.011603} {\bibfield  {journal} {\bibinfo  {journal} {Phys. Rev. A}\ }\textbf {\bibinfo {volume} {96}},\ \bibinfo {pages} {011603} (\bibinfo {year} {2017})}\BibitemShut {NoStop}%
\bibitem [{\citenamefont {Chen}\ \emph {et~al.}(2018{\natexlab{a}})\citenamefont {Chen}, \citenamefont {Lin}, \citenamefont {Chen}, \citenamefont {Chiu}, \citenamefont {Wang}, \citenamefont {Chen}, \citenamefont {Huang}, \citenamefont {Yip}, \citenamefont {Kawaguchi},\ and\ \citenamefont {Lin}}]{chen2018spin}%
  \BibitemOpen
  \bibfield  {author} {\bibinfo {author} {\bibfnamefont {H.-R.}\ \bibnamefont {Chen}}, \bibinfo {author} {\bibfnamefont {K.-Y.}\ \bibnamefont {Lin}}, \bibinfo {author} {\bibfnamefont {P.-K.}\ \bibnamefont {Chen}}, \bibinfo {author} {\bibfnamefont {N.-C.}\ \bibnamefont {Chiu}}, \bibinfo {author} {\bibfnamefont {J.-B.}\ \bibnamefont {Wang}}, \bibinfo {author} {\bibfnamefont {C.-A.}\ \bibnamefont {Chen}}, \bibinfo {author} {\bibfnamefont {P.}~\bibnamefont {Huang}}, \bibinfo {author} {\bibfnamefont {S.-K.}\ \bibnamefont {Yip}}, \bibinfo {author} {\bibfnamefont {Y.}~\bibnamefont {Kawaguchi}},\ and\ \bibinfo {author} {\bibfnamefont {Y.-J.}\ \bibnamefont {Lin}},\ }\bibfield  {title} {\bibinfo {title} {Spin--orbital-angular-momentum coupled bose-einstein condensates},\ }\href {https://doi.org/10.1103/PhysRevLett.121.113204} {\bibfield  {journal} {\bibinfo  {journal} {Phys. Rev. Lett.}\ }\textbf {\bibinfo {volume} {121}},\ \bibinfo {pages} {113204} (\bibinfo {year} {2018}{\natexlab{a}})}\BibitemShut {NoStop}%
\bibitem [{\citenamefont {Chen}\ \emph {et~al.}(2018{\natexlab{b}})\citenamefont {Chen}, \citenamefont {Liu}, \citenamefont {Tsai}, \citenamefont {Chiu}, \citenamefont {Kawaguchi}, \citenamefont {Yip}, \citenamefont {Chang},\ and\ \citenamefont {Lin}}]{chen2018rotating}%
  \BibitemOpen
  \bibfield  {author} {\bibinfo {author} {\bibfnamefont {P.-K.}\ \bibnamefont {Chen}}, \bibinfo {author} {\bibfnamefont {L.-R.}\ \bibnamefont {Liu}}, \bibinfo {author} {\bibfnamefont {M.-J.}\ \bibnamefont {Tsai}}, \bibinfo {author} {\bibfnamefont {N.-C.}\ \bibnamefont {Chiu}}, \bibinfo {author} {\bibfnamefont {Y.}~\bibnamefont {Kawaguchi}}, \bibinfo {author} {\bibfnamefont {S.-K.}\ \bibnamefont {Yip}}, \bibinfo {author} {\bibfnamefont {M.-S.}\ \bibnamefont {Chang}},\ and\ \bibinfo {author} {\bibfnamefont {Y.-J.}\ \bibnamefont {Lin}},\ }\bibfield  {title} {\bibinfo {title} {Rotating atomic quantum gases with light-induced azimuthal gauge potentials and the observation of the hess-fairbank effect},\ }\href {https://doi.org/10.1103/PhysRevLett.121.250401} {\bibfield  {journal} {\bibinfo  {journal} {Phys. Rev. Lett.}\ }\textbf {\bibinfo {volume} {121}},\ \bibinfo {pages} {250401} (\bibinfo {year} {2018}{\natexlab{b}})}\BibitemShut {NoStop}%
\bibitem [{\citenamefont {Zhang}\ \emph {et~al.}(2019)\citenamefont {Zhang}, \citenamefont {Gao}, \citenamefont {Zou}, \citenamefont {Kong}, \citenamefont {Li}, \citenamefont {Shen}, \citenamefont {Chen}, \citenamefont {Peng}, \citenamefont {Zhan}, \citenamefont {Pu},\ and\ \citenamefont {Jiang}}]{zhang2019ground}%
  \BibitemOpen
  \bibfield  {author} {\bibinfo {author} {\bibfnamefont {D.}~\bibnamefont {Zhang}}, \bibinfo {author} {\bibfnamefont {T.}~\bibnamefont {Gao}}, \bibinfo {author} {\bibfnamefont {P.}~\bibnamefont {Zou}}, \bibinfo {author} {\bibfnamefont {L.}~\bibnamefont {Kong}}, \bibinfo {author} {\bibfnamefont {R.}~\bibnamefont {Li}}, \bibinfo {author} {\bibfnamefont {X.}~\bibnamefont {Shen}}, \bibinfo {author} {\bibfnamefont {X.-L.}\ \bibnamefont {Chen}}, \bibinfo {author} {\bibfnamefont {S.-G.}\ \bibnamefont {Peng}}, \bibinfo {author} {\bibfnamefont {M.}~\bibnamefont {Zhan}}, \bibinfo {author} {\bibfnamefont {H.}~\bibnamefont {Pu}},\ and\ \bibinfo {author} {\bibfnamefont {K.}~\bibnamefont {Jiang}},\ }\bibfield  {title} {\bibinfo {title} {Ground-state phase diagram of a spin-orbital-angular-momentum coupled bose-einstein condensate},\ }\href {https://doi.org/10.1103/PhysRevLett.122.110402} {\bibfield  {journal} {\bibinfo  {journal} {Phys. Rev. Lett.}\ }\textbf {\bibinfo {volume} {122}},\ \bibinfo {pages} {110402} (\bibinfo
  {year} {2019})}\BibitemShut {NoStop}%
\bibitem [{\citenamefont {Peng}\ \emph {et~al.}(2022)\citenamefont {Peng}, \citenamefont {Jiang}, \citenamefont {Chen}, \citenamefont {Chen}, \citenamefont {Zou},\ and\ \citenamefont {He}}]{peng2022spin}%
  \BibitemOpen
  \bibfield  {author} {\bibinfo {author} {\bibfnamefont {S.-G.}\ \bibnamefont {Peng}}, \bibinfo {author} {\bibfnamefont {K.}~\bibnamefont {Jiang}}, \bibinfo {author} {\bibfnamefont {X.-L.}\ \bibnamefont {Chen}}, \bibinfo {author} {\bibfnamefont {K.-J.}\ \bibnamefont {Chen}}, \bibinfo {author} {\bibfnamefont {P.}~\bibnamefont {Zou}},\ and\ \bibinfo {author} {\bibfnamefont {L.}~\bibnamefont {He}},\ }\bibfield  {title} {\bibinfo {title} {Spin-orbital-angular-momentum-coupled quantum gases},\ }\href {https://doi.org/10.1007/s43673-022-00069-w} {\bibfield  {journal} {\bibinfo  {journal} {AAPPS Bulletin}\ }\textbf {\bibinfo {volume} {32}},\ \bibinfo {pages} {36} (\bibinfo {year} {2022})}\BibitemShut {NoStop}%
\bibitem [{\citenamefont {Chen}\ \emph {et~al.}(2020{\natexlab{a}})\citenamefont {Chen}, \citenamefont {Peng}, \citenamefont {Zou}, \citenamefont {Liu},\ and\ \citenamefont {Hu}}]{chen2020angular}%
  \BibitemOpen
  \bibfield  {author} {\bibinfo {author} {\bibfnamefont {X.-L.}\ \bibnamefont {Chen}}, \bibinfo {author} {\bibfnamefont {S.-G.}\ \bibnamefont {Peng}}, \bibinfo {author} {\bibfnamefont {P.}~\bibnamefont {Zou}}, \bibinfo {author} {\bibfnamefont {X.-J.}\ \bibnamefont {Liu}},\ and\ \bibinfo {author} {\bibfnamefont {H.}~\bibnamefont {Hu}},\ }\bibfield  {title} {\bibinfo {title} {Angular stripe phase in spin-orbital-angular-momentum coupled bose condensates},\ }\href {https://doi.org/10.1103/PhysRevResearch.2.033152} {\bibfield  {journal} {\bibinfo  {journal} {Phys. Rev. Res.}\ }\textbf {\bibinfo {volume} {2}},\ \bibinfo {pages} {033152} (\bibinfo {year} {2020}{\natexlab{a}})}\BibitemShut {NoStop}%
\bibitem [{\citenamefont {Chen}\ \emph {et~al.}(2020{\natexlab{b}})\citenamefont {Chen}, \citenamefont {Wu}, \citenamefont {Peng}, \citenamefont {Yi},\ and\ \citenamefont {He}}]{chen2020generating}%
  \BibitemOpen
  \bibfield  {author} {\bibinfo {author} {\bibfnamefont {K.-J.}\ \bibnamefont {Chen}}, \bibinfo {author} {\bibfnamefont {F.}~\bibnamefont {Wu}}, \bibinfo {author} {\bibfnamefont {S.-G.}\ \bibnamefont {Peng}}, \bibinfo {author} {\bibfnamefont {W.}~\bibnamefont {Yi}},\ and\ \bibinfo {author} {\bibfnamefont {L.}~\bibnamefont {He}},\ }\bibfield  {title} {\bibinfo {title} {Generating giant vortex in a fermi superfluid via spin-orbital-angular-momentum coupling},\ }\href {https://doi.org/10.1103/PhysRevLett.125.260407} {\bibfield  {journal} {\bibinfo  {journal} {Phys. Rev. Lett.}\ }\textbf {\bibinfo {volume} {125}},\ \bibinfo {pages} {260407} (\bibinfo {year} {2020}{\natexlab{b}})}\BibitemShut {NoStop}%
\bibitem [{\citenamefont {Wang}\ \emph {et~al.}(2021{\natexlab{b}})\citenamefont {Wang}, \citenamefont {Ji}, \citenamefont {Sun},\ and\ \citenamefont {Li}}]{wang2021exotic}%
  \BibitemOpen
  \bibfield  {author} {\bibinfo {author} {\bibfnamefont {L.-L.}\ \bibnamefont {Wang}}, \bibinfo {author} {\bibfnamefont {A.-C.}\ \bibnamefont {Ji}}, \bibinfo {author} {\bibfnamefont {Q.}~\bibnamefont {Sun}},\ and\ \bibinfo {author} {\bibfnamefont {J.}~\bibnamefont {Li}},\ }\bibfield  {title} {\bibinfo {title} {Exotic vortex states with discrete rotational symmetry in atomic fermi gases with spin-orbital--angular-momentum coupling},\ }\href {https://doi.org/10.1103/PhysRevLett.126.193401} {\bibfield  {journal} {\bibinfo  {journal} {Phys. Rev. Lett.}\ }\textbf {\bibinfo {volume} {126}},\ \bibinfo {pages} {193401} (\bibinfo {year} {2021}{\natexlab{b}})}\BibitemShut {NoStop}%
\bibitem [{\citenamefont {Duan}\ \emph {et~al.}(2020)\citenamefont {Duan}, \citenamefont {Bidasyuk},\ and\ \citenamefont {Surzhykov}}]{duan2020symmetry}%
  \BibitemOpen
  \bibfield  {author} {\bibinfo {author} {\bibfnamefont {Y.}~\bibnamefont {Duan}}, \bibinfo {author} {\bibfnamefont {Y.~M.}\ \bibnamefont {Bidasyuk}},\ and\ \bibinfo {author} {\bibfnamefont {A.}~\bibnamefont {Surzhykov}},\ }\bibfield  {title} {\bibinfo {title} {Symmetry breaking and phase transitions in bose-einstein condensates with spin--orbital-angular-momentum coupling},\ }\href {https://doi.org/10.1103/PhysRevA.102.063328} {\bibfield  {journal} {\bibinfo  {journal} {Phys. Rev. A}\ }\textbf {\bibinfo {volume} {102}},\ \bibinfo {pages} {063328} (\bibinfo {year} {2020})}\BibitemShut {NoStop}%
\bibitem [{\citenamefont {Bidasyuk}\ \emph {et~al.}(2022)\citenamefont {Bidasyuk}, \citenamefont {Kovtunenko},\ and\ \citenamefont {Prikhodko}}]{bidasyuk2022fine}%
  \BibitemOpen
  \bibfield  {author} {\bibinfo {author} {\bibfnamefont {Y.~M.}\ \bibnamefont {Bidasyuk}}, \bibinfo {author} {\bibfnamefont {K.~S.}\ \bibnamefont {Kovtunenko}},\ and\ \bibinfo {author} {\bibfnamefont {O.~O.}\ \bibnamefont {Prikhodko}},\ }\bibfield  {title} {\bibinfo {title} {Fine structure of the stripe phase in ring-shaped bose-einstein condensates with spin-orbital-angular-momentum coupling},\ }\href {https://doi.org/10.1103/PhysRevA.105.023320} {\bibfield  {journal} {\bibinfo  {journal} {Phys. Rev. A}\ }\textbf {\bibinfo {volume} {105}},\ \bibinfo {pages} {023320} (\bibinfo {year} {2022})}\BibitemShut {NoStop}%
\bibitem [{\citenamefont {Chen}\ \emph {et~al.}(2022)\citenamefont {Chen}, \citenamefont {Wu}, \citenamefont {He},\ and\ \citenamefont {Yi}}]{chen2022angular}%
  \BibitemOpen
  \bibfield  {author} {\bibinfo {author} {\bibfnamefont {K.-J.}\ \bibnamefont {Chen}}, \bibinfo {author} {\bibfnamefont {F.}~\bibnamefont {Wu}}, \bibinfo {author} {\bibfnamefont {L.}~\bibnamefont {He}},\ and\ \bibinfo {author} {\bibfnamefont {W.}~\bibnamefont {Yi}},\ }\bibfield  {title} {\bibinfo {title} {Angular topological superfluid and topological vortex in an ultracold fermi gas},\ }\href {https://doi.org/10.1103/PhysRevResearch.4.033023} {\bibfield  {journal} {\bibinfo  {journal} {Phys. Rev. Res.}\ }\textbf {\bibinfo {volume} {4}},\ \bibinfo {pages} {033023} (\bibinfo {year} {2022})}\BibitemShut {NoStop}%
\bibitem [{\citenamefont {Cao}\ \emph {et~al.}(2022)\citenamefont {Cao}, \citenamefont {Han}, \citenamefont {Wu}, \citenamefont {Yuan}, \citenamefont {He},\ and\ \citenamefont {Li}}]{cao2022quantum}%
  \BibitemOpen
  \bibfield  {author} {\bibinfo {author} {\bibfnamefont {R.}~\bibnamefont {Cao}}, \bibinfo {author} {\bibfnamefont {J.}~\bibnamefont {Han}}, \bibinfo {author} {\bibfnamefont {J.}~\bibnamefont {Wu}}, \bibinfo {author} {\bibfnamefont {J.}~\bibnamefont {Yuan}}, \bibinfo {author} {\bibfnamefont {L.}~\bibnamefont {He}},\ and\ \bibinfo {author} {\bibfnamefont {Y.}~\bibnamefont {Li}},\ }\bibfield  {title} {\bibinfo {title} {Quantum phases of spin-orbital-angular-momentum--coupled bosonic gases in optical lattices},\ }\href {https://doi.org/10.1103/PhysRevA.105.063308} {\bibfield  {journal} {\bibinfo  {journal} {Phys. Rev. A}\ }\textbf {\bibinfo {volume} {105}},\ \bibinfo {pages} {063308} (\bibinfo {year} {2022})}\BibitemShut {NoStop}%
\bibitem [{\citenamefont {Han}\ \emph {et~al.}(2022)\citenamefont {Han}, \citenamefont {Peng}, \citenamefont {Chen},\ and\ \citenamefont {Yi}}]{han2022molecular}%
  \BibitemOpen
  \bibfield  {author} {\bibinfo {author} {\bibfnamefont {Y.}~\bibnamefont {Han}}, \bibinfo {author} {\bibfnamefont {S.-G.}\ \bibnamefont {Peng}}, \bibinfo {author} {\bibfnamefont {K.-J.}\ \bibnamefont {Chen}},\ and\ \bibinfo {author} {\bibfnamefont {W.}~\bibnamefont {Yi}},\ }\bibfield  {title} {\bibinfo {title} {Molecular state in a spin--orbital-angular-momentum coupled fermi gas},\ }\href {https://doi.org/10.1103/PhysRevA.106.043302} {\bibfield  {journal} {\bibinfo  {journal} {Phys. Rev. A}\ }\textbf {\bibinfo {volume} {106}},\ \bibinfo {pages} {043302} (\bibinfo {year} {2022})}\BibitemShut {NoStop}%
\bibitem [{\citenamefont {Busch}\ \emph {et~al.}(1998)\citenamefont {Busch}, \citenamefont {Englert}, \citenamefont {Rza{\.z}ewski},\ and\ \citenamefont {Wilkens}}]{busch1998two}%
  \BibitemOpen
  \bibfield  {author} {\bibinfo {author} {\bibfnamefont {T.}~\bibnamefont {Busch}}, \bibinfo {author} {\bibfnamefont {B.-G.}\ \bibnamefont {Englert}}, \bibinfo {author} {\bibfnamefont {K.}~\bibnamefont {Rza{\.z}ewski}},\ and\ \bibinfo {author} {\bibfnamefont {M.}~\bibnamefont {Wilkens}},\ }\bibfield  {title} {\bibinfo {title} {Two cold atoms in a harmonic trap},\ }\href {https://doi.org/10.1023/A:1018705520999} {\bibfield  {journal} {\bibinfo  {journal} {Foundations of Physics}\ }\textbf {\bibinfo {volume} {28}},\ \bibinfo {pages} {549} (\bibinfo {year} {1998})}\BibitemShut {NoStop}%
\bibitem [{\citenamefont {Liu}\ \emph {et~al.}(2010)\citenamefont {Liu}, \citenamefont {Hu},\ and\ \citenamefont {Drummond}}]{liu2010exact}%
  \BibitemOpen
  \bibfield  {author} {\bibinfo {author} {\bibfnamefont {X.-J.}\ \bibnamefont {Liu}}, \bibinfo {author} {\bibfnamefont {H.}~\bibnamefont {Hu}},\ and\ \bibinfo {author} {\bibfnamefont {P.~D.}\ \bibnamefont {Drummond}},\ }\bibfield  {title} {\bibinfo {title} {Exact few-body results for strongly correlated quantum gases in two dimensions},\ }\href {https://doi.org/10.1103/PhysRevB.82.054524} {\bibfield  {journal} {\bibinfo  {journal} {Phys. Rev. B}\ }\textbf {\bibinfo {volume} {82}},\ \bibinfo {pages} {054524} (\bibinfo {year} {2010})}\BibitemShut {NoStop}%
\bibitem [{\citenamefont {Peng}\ \emph {et~al.}(2011{\natexlab{a}})\citenamefont {Peng}, \citenamefont {Li}, \citenamefont {Drummond},\ and\ \citenamefont {Liu}}]{peng2011high}%
  \BibitemOpen
  \bibfield  {author} {\bibinfo {author} {\bibfnamefont {S.-G.}\ \bibnamefont {Peng}}, \bibinfo {author} {\bibfnamefont {S.-Q.}\ \bibnamefont {Li}}, \bibinfo {author} {\bibfnamefont {P.~D.}\ \bibnamefont {Drummond}},\ and\ \bibinfo {author} {\bibfnamefont {X.-J.}\ \bibnamefont {Liu}},\ }\bibfield  {title} {\bibinfo {title} {High-temperature thermodynamics of strongly interacting $s$-wave and $p$-wave fermi gases in a harmonic trap},\ }\href {https://doi.org/10.1103/PhysRevA.83.063618} {\bibfield  {journal} {\bibinfo  {journal} {Phys. Rev. A}\ }\textbf {\bibinfo {volume} {83}},\ \bibinfo {pages} {063618} (\bibinfo {year} {2011}{\natexlab{a}})}\BibitemShut {NoStop}%
\bibitem [{\citenamefont {Peng}\ \emph {et~al.}(2011{\natexlab{b}})\citenamefont {Peng}, \citenamefont {Liu}, \citenamefont {Hu},\ and\ \citenamefont {Li}}]{peng2011non}%
  \BibitemOpen
  \bibfield  {author} {\bibinfo {author} {\bibfnamefont {S.-G.}\ \bibnamefont {Peng}}, \bibinfo {author} {\bibfnamefont {X.-J.}\ \bibnamefont {Liu}}, \bibinfo {author} {\bibfnamefont {H.}~\bibnamefont {Hu}},\ and\ \bibinfo {author} {\bibfnamefont {S.-Q.}\ \bibnamefont {Li}},\ }\bibfield  {title} {\bibinfo {title} {Non-universal thermodynamics of a strongly interacting inhomogeneous fermi gas using the quantum virial expansion},\ }\href {https://doi.org/https://doi.org/10.1016/j.physleta.2011.06.045} {\bibfield  {journal} {\bibinfo  {journal} {Physics Letters A}\ }\textbf {\bibinfo {volume} {375}},\ \bibinfo {pages} {2979} (\bibinfo {year} {2011}{\natexlab{b}})}\BibitemShut {NoStop}%
\end{thebibliography}%

\appendix
\section{Decomposing of \texorpdfstring{$V\left(r_{12}\right)$}{Lg} in the angular basis\label{sec:AppA}}

Let us expand the interaction potential $V\left(r_{12}\right)$ in the basis of the angular eigenstates of $\hat{l}_{1z}$ and $\hat{l}_{2z}$. To this end, the two-body potential can be written as
\begin{equation}
V\left(r_{12}\right)=\int d{\bf r}V\left(r\right)\delta\left({\bf r}_{1}-{\bf r}_{2}-{\bf r}\right).
\end{equation}
Since we have
\begin{equation}
\delta\left({\bf r}\right)=\frac{1}{\left(2\pi\right)^{2}}\int d{\bf k}e^{i{\bf k}\cdot{\bf r}},
\end{equation}
it yields
\begin{equation} \label{eq:V_r12}
V\left(r_{12}\right)=\int d{\bf r}\frac{V\left(r\right)}{\left(2\pi\right)^{2}}\int d{\bf k}e^{i{\bf k}\cdot\left({\bf r}_{1}-{\bf r}_{2}-{\bf r}\right)}.
\end{equation}
By using the 2D plane wave expansion, 
\begin{equation}
e^{i{\bf k}\cdot{\bf r}}=\sum_{l=-\infty}^{\infty}i^{l}J_{l}\left(kr\right)e^{il\left(\varphi_{{\bf k}}-\varphi_{{\bf r}}\right)},
\end{equation}
we find
\begin{eqnarray}
V\left(r_{12}\right) & = & \int d{\bf r}\frac{V\left(r\right)}{\left(2\pi\right)^{2}}\int d{\bf k}e^{i{\bf k}\cdot\left({\bf r}_{1}-{\bf r}_{2}\right)}\cdot e^{-i{\bf k}\cdot{\bf r}}\nonumber \\
 & = & \frac{1}{2\pi}\int_{0}^{\infty}V\left(r\right)rdr\int e^{i{\bf k}\cdot\left({\bf r}_{1}-{\bf r}_{2}\right)}J_{0}\left(kr\right)kdkd\varphi_{{\bf k}}\nonumber \\
 & = & \sum_{l=-\infty}^{\infty}V_{l}\left(r_{1},r_{2}\right)e^{-il\left(\varphi_{1}-\varphi_{2}\right)}
\end{eqnarray}
with
\begin{equation}
V_{l}\left(r_{1},r_{2}\right)=\int_{0}^{\infty}rdrV\left(r\right)\int_{0}^{\infty}kdkJ_{0}\left(kr\right)J_{l}\left(kr_{1}\right)J_{l}\left(kr_{2}\right).
\end{equation}

\section{The explicit form of \texorpdfstring{$\boldsymbol{\mathcal{V}}$}{Lg} matrix}\label{app:V-matrix}

The explicit forms of four matrix 
 elements $\boldsymbol{\mathcal{V}}$ in Eq.~\eqref{eq:2bSecularEq} are given by
\begin{widetext}
\begin{subequations}
    \begin{eqnarray}
\mathcal{V}_{\left(k_{1}k_{2}l\right)\left(k_{1}^{\prime}k_{2}^{\prime}l^{\prime}\right)}^{\left(\uparrow\downarrow\right)\left(\uparrow\downarrow\right)} & \equiv & \iint\left[u^{k_{1}}_{l,\uparrow}\left(r_{1}\right)u^{k_{2}}_{-l,\downarrow}\left(r_{2}\right)\frac{V_{l^{\prime}-l}\left(r_{1},r_{2}\right)}{2}u^{k_{1}^{\prime}}_{l^{\prime},\uparrow}\left(r_{1}\right)u^{k_{2}^{\prime}}_{-l^{\prime},\downarrow}\left(r_{2}\right)\right]r_{1}r_{2}dr_{1}dr_{2},\\
\mathcal{V}_{\left(k_{1}k_{2}l\right)\left(k_{1}^{\prime}k_{2}^{\prime}l^{\prime}\right)}^{\left(\uparrow\downarrow\right)\left(\downarrow\uparrow\right)} & \equiv & \iint\left[u^{k_{1}}_{l,\uparrow}\left(r_{1}\right)u^{k_{2}}_{-l,\downarrow}\left(r_{2}\right)\frac{V_{l^{\prime}-l}\left(r_{1},r_{2}\right)}{2}u^{k_{1}^{\prime}}_{l^{\prime},\downarrow}\left(r_{1}\right)u^{k_{2}^{\prime}}_{-l^{\prime},\uparrow}\left(r_{2}\right)\right]r_{1}r_{2}dr_{1}dr_{2},\\
\mathcal{V}_{\left(k_{1}k_{2}l\right)\left(k_{1}^{\prime}k_{2}^{\prime}l^{\prime}\right)}^{\left(\downarrow\uparrow\right)\left(\downarrow\uparrow\right)} & \equiv & \iint\left[u^{k_{1}}_{l,\downarrow}\left(r_{1}\right)u^{k_{2}}_{-l,\uparrow}\left(r_{2}\right)\frac{V_{l^{\prime}-l}\left(r_{1},r_{2}\right)}{2}u^{k_{1}^{\prime}}_{l^{\prime},\downarrow}\left(r_{1}\right)u^{k_{2}^{\prime}}_{-l^{\prime},\uparrow}\left(r_{2}\right)\right]r_{1}r_{2}dr_{1}dr_{2},\\
\mathcal{V}_{\left(k_{1}k_{2}l\right)\left(k_{1}^{\prime}k_{2}^{\prime}l^{\prime}\right)}^{\left(\downarrow\uparrow\right)\left(\uparrow\downarrow\right)} & \equiv & \iint\left[u^{k_{1}}_{l,\downarrow}\left(r_{1}\right)u^{k_{2}}_{-l,\uparrow}\left(r_{2}\right)\frac{V_{l^{\prime}-l}\left(r_{1},r_{2}\right)}{2}u^{k_{1}^{\prime}}_{l^{\prime},\uparrow}\left(r_{1}\right)u^{k_{2}^{\prime}}_{-l^{\prime},\downarrow}\left(r_{2}\right)\right]r_{1}r_{2}dr_{1}dr_{2}.
\end{eqnarray}
\end{subequations}
\end{widetext}

\section{Scattering parameters for a SSW potential} \label{app:scattering_length}

Let us consider the scattering problem of two atoms interacting with a SSW potential in 2D. The relative motion of two atoms is described by the following equation
\begin{equation}
\left[-\frac{\hbar^{2}}{m}\nabla^{2}+V\left({\bf r}\right)\right]\psi\left({\bf r}\right)=E\psi\left({\bf r}\right)
\end{equation}
with
\begin{equation}
V\left({\bf r}\right)=\begin{cases}
-V_{0}, & 0\le r\le\epsilon,\\
0, & r>\epsilon,
\end{cases}
\end{equation}
and $E>0$ for a scattering problem. The angular momentum is a good quantum number: thus, the wave function is written as $\psi\left({\bf r}\right)=u_{l}\left(r\right)e^{il\varphi}/\sqrt{2\pi}$, which yields the radial equation:
\begin{equation}
\left[\frac{1}{r}\frac{\partial}{\partial r}r\frac{\partial}{\partial r}-\frac{l^{2}}{r^{2}}+k^{2}-\frac{mV\left(r\right)}{\hbar^{2}}\right]u_{l}\left(r\right)=0
\end{equation}
with $k^{2}=mE/\hbar^{2}$. For the $s$-wave scattering, i.e., $l=0$, the solution takes the form of
\begin{equation}
u_{0}\left(r\right)=\begin{cases}
cJ_{0}\left(Gr\right), & 0\le r\le\epsilon,\\
\cot\delta_{0}\cdot J_{0}\left(kr\right)-N_{0}\left(kr\right), & r>\epsilon,
\end{cases}
\end{equation}
where $\delta_{0}$ is the $s$-wave scattering phase shift, $c$ is a normalization parameter, and $G^{2}=k^{2}+mV_{0}/\hbar^{2}$. Here, $J_{\nu}\left(\cdot\right)$ and $N_{\nu}\left(\cdot\right)$ are Bessel functions of the first and second kinds. By using the continuity condition of the wave function and its first-order derivative at $r=\epsilon$, we easily obtain the scattering phase shift
\begin{equation}
\cot\delta_{0}=\frac{\left(k\epsilon\right)J_{0}\left(G\epsilon\right)N_{1}\left(k\epsilon\right)-\left(G\epsilon\right)J_{1}\left(G\epsilon\right)N_{0}\left(k\epsilon\right)}{\left(k\epsilon\right)J_{0}\left(G\epsilon\right)J_{1}\left(k\epsilon\right)-\left(G\epsilon\right)J_{1}\left(G\epsilon\right)J_{0}\left(k\epsilon\right)}.
\end{equation}
Expanding $\cot\delta_{0}$ at small $k$, we obtain the effective-range expansion of the scattering phase shift in 2D, i.e.,
\begin{equation}
\cot\delta_{0}=\frac{2}{\pi}\ln\left(ka_{2D}\right)+O\left(k^{2}\right),
\end{equation}
with the 2D $s$-wave scattering length: 
\begin{equation}
\ln\frac{a_{2D}}{\epsilon}=\frac{J_{0}\left(\sqrt{\tilde{V}_{0}}\right)}{\sqrt{\tilde{V}_{0}}J_{1}\left(\sqrt{\tilde{V}_{0}}\right)}+\gamma_{E}-\ln2,
\end{equation}
with Euler gamma constant $\gamma_{E}\approx0.577216$ and $\tilde{V}_{0}\equiv V_{0}/\left(\hbar^{2}/m\epsilon^{2}\right)$, or in a form of
\begin{equation}
\ln\frac{a_{2d}}{a_\mathrm{ho}}=\frac{J_{0}\left(\sqrt{m\epsilon^{2}V_{0}/\hbar^{2}}\right)}{\sqrt{m\epsilon^{2}V_{0}/\hbar^{2}}J_{1}\left(\sqrt{m\epsilon^{2}V_{0}/\hbar^{2}}\right)}-\ln\frac{2a_\mathrm{ho}}{\epsilon e^{\gamma_{E}}}. 
\end{equation} 
Explicitly, the relationship between this introduced scatter length $\ln{(a_{2d}/a_\mathrm{ho})}$ and the two-body interaction strength $V_0$ can be seen in the right plot of Fig.~\ref{figure3_En_lna2dB_n0}.

\end{document}